\documentclass[%
reprint,
superscriptaddress,
amsmath,amssymb,
aps,
prapplied,
]{revtex4-2}

\usepackage{graphicx}
\usepackage{dcolumn}
\usepackage{bm}
\usepackage{hyperref}

\usepackage{lipsum}
\usepackage[]{siunitx}
\usepackage[version=4]{mhchem}
\usepackage{textcomp} 
\usepackage{enumitem}

\makeatletter
\def\maketitle{
	\@author@finish
	\title@column\titleblock@produce
	\suppressfloats[t]}
\makeatother 

\begin{document}
	
	\preprint{APS/123-QED}
	
	\title{Mid-Infrared Integrated Acousto-Optic Phase Modulation on GaAs/AlGaAs}
	
	\author{Thomas~Gérodou}\email{thomas.gerodou@universite-paris-saclay.fr}
	\author{Mathilde~Urbain}
	\author{Luca~Lucia}
	\author{Etienne~Herth}
	\author{Jean-René~Coudevylle}
	\author{Martina~Morassi}
	\author{Aristide~Lemaître}
	\affiliation{Université Paris-Saclay, CNRS, Centre de Nanosciences et de Nanotechnologies, 91120 Palaiseau, France}
	\author{Damien~Teyssieux}
	\affiliation{Université de Franche-Comté, ENSMM, CNRS, Institut FEMTO-ST, 25030 Besançon, France}
	\author{Raffaele~Colombelli}
	\author{Adel~Bousseksou}
	\author{Mathieu~Jeannin}\email{mathieu.jeannin@cnrs.fr}
	\affiliation{Université Paris-Saclay, CNRS, Centre de Nanosciences et de Nanotechnologies, 91120 Palaiseau, France}

	\date{\today}
	
	\begin{abstract}
We demonstrate a mid-infrared ($\lambda\in[$7.5--10]~\si{\micro m}) integrated acousto-optic phase modulator operating at \qty{460}{MHz} based on a GaAs/AlGaAs heterostructure. The combination of broad mid-infrared transparency and piezoelectric properties of GaAs based structures allows monolithic integration of all the device functions (electro-acoustic transduction and acousto-optic interaction) on a single material platform. 
Addressing the challenge presented by comparable mid-infrared and acoustic wavelengths, simultaneous propagation and confinement of surface acoustic waves and guided light is achieved using a deep-etched sub-wavelength grating.
We develop a comprehensive and quantitative simulation framework to design the modulator devices, characterize their acoustic properties in a heterodyne vibrometer and perform a full, quantitative analysis of the device performance by inserting them in an integrated Mach-Zehnder interferometer. 
Remarkable agreement with simulations validates the proposed design flow, and opens the way to improved integrated acousto-optic devices operating in the longer end of the mid-infrared.
	\end{abstract}
	
	\maketitle
	
	
	\section{\label{sec:level1}Introduction}

The mid-infrared spectral region (mid-IR, {${\lambda\in}[$3--12]~\si{\micro m}}) has seen great progress in the past decade, with significant applications in particular related to molecular spectroscopy. While mid-IR sources (quantum or interband cascade lasers, QCL or ICL) and detectors are now commercially available, the implementation of more advanced mid-IR opto-electronic functions run far behind their visible and near-IR counterparts.
Among desirable operations, the ability to control the phase or intensity of a laser beam is of paramount importance, with requirements of high efficiency and operating speeds.

Since the demonstration of the QCL, unipolar devices relying on intersubband transitions played a great role in the development of mid-IR optoelectronics. Novel device concepts emerged recently, employing cavity electrodynamics designs to enhance light-matter interaction, enabling demonstrations of ultra-high bandwidth detectors \cite{hakl2021ultrafastquantumwell, lin2023realtimechirpedpulse}, modulators based on the quantum-confined Stark effect \cite{dely202210gbit, bonazzi2024metamaterialunipolar} or switching of the light-matter coupling strength \cite{pirotta2021fastamplitude, malerba2024ultrafast10, ngo2026midirsingle}. Interestingly, this latter concept has also been proposed to demonstrate non-linear optical devices such as power limiters or saturable absorbers \cite{mann2021ultrafastoptical, jeannin2021unifieddescription, jeannin2023lowintensity}. 
One particular application of many of these devices is the generation of stable, compact and efficient frequency combs with the potential to replace QCL-based combs by more efficient systems for dual comb spectroscopy, whether using electro-optic comb formation \cite{ngo2026midirsingle} or passive laser mode-locking using saturable absorbers \cite{jeannin2023lowintensity}.

Besides the use of microcavities, another promising route to increase light-matter interaction in optoelectronic devices is their integration in waveguides on photonic chips. Benefiting from the extensive development of integrated photonics driven by telecommunication applications, recent interest in mid-IR photonic integration has lead to impressive advances on the path towards downsizing current bulky tabletop optical systems into on-chip circuits. 
Material platforms based on Ge and Si have already shown great progress building on the mature silicon photonics environment \cite{marris-morini2026germaniumbasedmidinfrared}, including demonstrations of passive circuits (waveguides, cavities) with record low propagation losses \cite{nguyen2024integratedpin, turpaud2024lowlosssige, turpaud2024tunableonchip, huertas-pedroche2025highspeedintegrated, perestjuk2025onemillion} and more advanced functions involving dispersion engineering \cite{bricout2026pulsegeneration}.
Motivated by the prospect of full monolithic integration from sources to detectors offered by \mbox{III-V} platforms  based on InP and GaAs \cite{jung2019homogeneousphotonic, wang2022monolithicintegration, karnik2025onchipwavelength}, integrated \mbox{III-V} mid-IR circuits emerged recently demonstrating competitive performance for passive structures \cite{zhang2022midinfraredmicroring, karnik2023highefficiencymidinfrared, montesinos-ballester2024lowlossburied, lucia2026midinfraredintegrated} with losses down to the dB/cm. 
Chalcogenide glasses are also being investigated as alternative candidates \cite{suk2025ultrahighqchalcogenide, ma2013lowlosschalcogenide}, albeit facing considerable fabrication challenges. 

Shortly after passive components, high-speed integrated amplitude modulators were demonstrated based on fast modulation of free-carriers absorption in SiGe \cite{huertas-pedroche2025highspeedintegrated}, and using intersubband transitions in III-V coupled quantum wells \cite{lucia2026fastmidinfrared, lucia2026activepassive}. Alongside free-space devices, integrated mid-IR devices are also racing towards micro-comb formation, either in an active form \cite{turpaud2024tunableonchip} or aiming at non-linear Kerr comb formation in low loss cavities \cite{perestjuk2025onemillion, lucia2026midinfraredintegrated}.

While the above developments include multiple approaches for fast amplitude modulation, integrated \textit{pure phase} modulation remains a challenge at high speeds (sub-GHz to GHz). Slower kHz-bandwidth thermo-optic devices have been reported \cite{huertas-pedroche2026integratedphase} on SiGe, but the main technological solutions for fast operation involve non-linear effects such as the electro-optic (EO) or acousto-optic (AO) effect. In short, EO modulators (EOM) allow a direct modulation of the refractive index with an externally applied voltage on the EO material, either though the Pockels of the Kerr effect. In contrast, AOM require the transduction of an applied voltage to an elastic wave through the inverse piezoelectric effect, followed by modulation of the refractive index supported by the photo-elastic interaction. 
An important difference between EOM and AOM is that EOM generally operate over a wide RF bandwidth (typ.\ DC to several GHz), while most of AOM are based on resonant excitation of acoustic waves in interdigital transducers (IDTs) that are inherently narrow band, with typical MHz bandwidths and sub-GHz to $\approx$\qty{10}{GHz} center frequencies. 
Both technologies require materials with high electro-optic coefficients (EOM) or a combination of large piezoelectric and photo-elastic coefficients (AOM). Interestingly, materials in state-of-the-art integrated devices are the same for EO and AO modulators, relying mainly on functional oxides such as lithium niobate or lithium tantalate. In that respect, short-wave infrared ($\lambda \approx 3$--\qty{4}{\micro m}) modulators have been demonstrated on thin film lithium niobate using both EO \cite{didier2026thinfilm, lee2026suspendedthinfilm} and AO \cite{xu2024unveilingefficient} effects.
Unfortunately, the presence of phonons in oxide groups always causes strong absorption in the longer wavelength region of the mid-IR, restricting the use of all widespread EO and AO materials to wavelengths below 5~\si{\micro m} \cite{lin2018midinfraredintegrated}.

With the goal of providing a monolithic platform at in the 8--\qty{10}{\micro m} range, \mbox{III-V} semiconductors appear as the main alternative. They provide broadband transparency, strong photo-elastic coefficients, and piezoelectricity albeit with small electromechanical coupling. 
GaAs in particular has well-documented, although relatively weak, piezoelectric properties allowing for piezoelectric generation of surface acoustic waves (SAW) and acousto-optic interaction within a single material. 
As the incorporation of Al in GaAs decreases the refractive index and increases the sound velocity, GaAs/AlGaAs heterostructures naturally form optical and acoustic waveguides with GaAs core and AlGaAs cladding for both optical and mechanical waves. 
While integrated acousto-optic modulation has been demonstrated on a GaAs/AlGaAs platform in the near-IR \cite{delima2006compactmachzehnder}, the longer mid-IR wavelength brings challenges beyond material transparency considerations. 
Given typical Rayleigh SAW velocity on GaAs of 2.86~km/s, SAW at 1~GHz have wavelengths comparable to the size of mid-IR waveguides, making destructive interference across the optical mode and complex acoustic behaviors possible. 
This makes a careful design of the interaction region geometry critical to the device, as evidenced in previous studies approaching the same type of regime from the different perspective of small SAW wavelengths \cite{tadesse2014subopticalwavelength, zhang2025onchip7, li2015nanophotoniccavity} (targeting high RF frequencies), instead of large optical wavelengths. 

In this work, we introduce a GaAs/AlGaAs platform for integrated acousto-optic phase modulation at the longer end of the mid-IR spectral region ($\lambda \approx 7.5-10$~\si{\micro m}). Focusing on the main challenge of engineering a good overlap between acoustic and optical waves of comparable wavelengths, we present a predictive and quantitative acousto-optic simulation scheme, coupled with a heterodyne vibrometry scheme for acoustic diagnostic. Using these tools, we demonstrate integrated phase modulation at optimized frequencies (section \ref{sec:design}) around \qty{460}{MHz}. The experimental variations in modulation efficiency and acoustic behavior with device geometry show remarkable agreement with simulations. This opens the way to realizing other acoustic functionalities in this wavelength range, leveraging the developed platform, design and characterization tools.

	\begin{figure*}[!t]
		\includegraphics{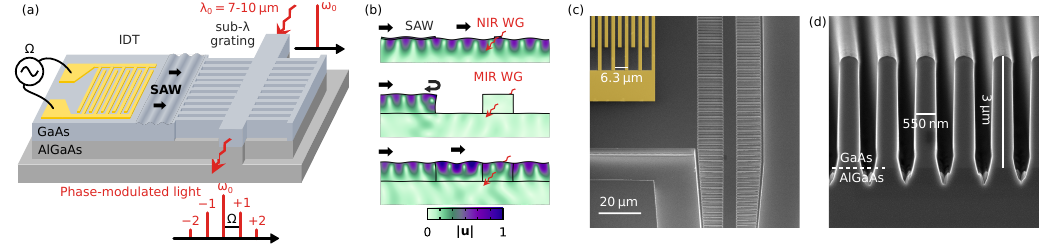}
		\caption{\label{fig:platform}GaAs/AlGaAs platform for mid-IR integrated acousto-optics. (a) Phase modulator scheme. A piezoelectric interdigital transducer (IDT) driven by an RF probe at frequency $\Omega$ generate surface acoustic waves on the GaAs substrate, which propagate through a sub-wavelength grating metamaterial onto a mid-IR optical waveguide. 
			The periodic mechanical strain induces through the acousto-optic interaction a modulation of the effective optical mode index that is uniform along the interaction length, resulting in a phase modulation of transmitted light. (b) Displacement profiles from finite element method simulations of \qty{459}{MHz} SAW on a GaAs/AlGaAs structure propagating towards an optical waveguide. Color indicates displacement magnitude at a fixed phase, normalized to the incident SAW amplitude. Top: SAW freely pass through a $1.4\times$\qty{0.3}{\micro m\squared} ridge guide suited for near-infrared light (NIR WG). Middle: SAW are reflected by deeper grooves and do not reach the larger $6.5\times$\qty{3}{\micro m\squared} guide compatible with target mid-IR wavelengths. Bottom: Transmission is restored using a \qty{1}{\micro m}-period grating as in (a). (c,d) SEM micrographs of a fabricated device. The IDT is highlighted in yellow. (d) shows a cross-section profile of the etched grating.}
	\end{figure*}
	\section{Device principle and material platform}

The acousto-optic phase modulator device sketched in Fig.~\ref{fig:platform}(a) is built on a GaAs/AlGaAs heterostructure. Rayleigh SAW are excited on the piezoelectric GaAs top surface by an IDT made of a series of patterned metal electrodes with periodicity $\Lambda$, driven by an RF signal at a frequency $\Omega = v_\text{SAW}/\Lambda$ where $v_\text{SAW}$ is the SAW phase velocity. The generated SAW then naturally propagate to an optical waveguide. The optical and acoustic waves are orthogonal, so the periodic mechanical strain induces a uniform modulation of the optical mode index along the interaction length through the acousto-optic interaction, resulting in a phase modulation of the transmitted light. The devices are oriented along the standard [011] crystal cut angle which hosts a Rayleigh pseudo-SAW with maximal piezoelectric coupling for (100)-GaAs \cite{hunt1986slownesssurface}.

In order to accommodate optical modes in the 7.5--\qty{10}{\micro m} region while keeping the optical field as close as possible to the surface, the GaAs waveguide core height is set to \qty{3}{\micro m}. Because of the large etch depth required to make such deep ridges, performing any lithography step after etching is impractical. We thus define the IDTs on the top surface in the very first steps, before etching.
In this situation defining simple grooves to isolate the waveguide, which is typically done at shorter wavelengths \cite{delima2006compactmachzehnder, tadesse2014subopticalwavelength, wan2022highlyefficient} (Fig.~\ref{fig:platform}(b), top) would induce strong reflections of the SAW in our case as the etch depth is comparable to the acoustic wavelength (Fig.~\ref{fig:platform}(b), middle).

Instead, we pattern a sub-optical-wavelength grating around the waveguide. The grating forms an effective index region with an index lower than the GaAs core, ensuring optical mode confinement. At the same time, acoustic transmission along the grating bridges is ensured through the mechanical continuity of the material (Fig.~\ref{fig:platform}(b), bottom). 
A sub-wavelength grating was selected over other classes of meta-materials because it offers broadband properties that do not rely on acoustic or optical bandgaps. More details about its optical properties can be found in Supplementary Materials \cite{seesupplementary}.

The AOM devices are fabricated on a GaAs/\ce{Al_{0.5}Ga_{0.5}As} hetero-structure grown by molecular beam epitaxy, composed of a \qty{3}{\micro m}-thick GaAs core on top of a \qty{7}{\micro m}-thick \ce{Al_{0.5}Ga_{0.5}As} cladding. We begin the fabrication process by defining the IDTs using electron-beam lithography (EBL), followed by metal evaporation and of a Ti-Al (\qty{10}{nm}--\qty{80}{nm}) layer and lift-off. We then perform a second lithography and Ti-Au evaporation step to thicken the electrical contact pads. 

Next, all optical structures are defined in a single EBL step, including waveguide contours in double-trench geometry, mesa regions hosting the IDTs and subwavelength gratings. We first deposit a \qty{1}{\micro m}-thick layer of SiN by plasma-enhanced chemical vapor deposition, topped by \qty{40}{nm} of Ge to ensure charge evacuation during exposure as well as correct e-beam focusing, and a thick \qty{1.1}{\micro m} spin-coated layer of CSAR 62 positive resist. We define the pattern on the resist by EBL exposure and development, then transfer it to the SiN in a \ce{CHF3}-based inductively coupled plasma reactive ion etching (ICP-RIE) step, defining a hard mask. After stripping the remaining resist chemically, we transfer the SiN mask into the  \mbox{III-V} heterostructure through a HBr-based ICP-RIE \cite{bouchoule2008anisotropicsmooth}. The remaining SiN mask is finally removed in a fluorine-based RIE.

Scanning electron microscopy (SEM) pictures of fabricated devices are shown in Fig.~\ref{fig:platform}(c,d), displaying in particular the full \qty{3}{\micro m} etch of the GaAs core layer in the grating region.
Particular care was taken to maintain vertical sidewalls throughout the process, including careful optimization of the resin profiles. 
Due to the higher etch rate in broader trench openings, the etch depth reaches \qty{6.5}{\micro m} in regions surrounding the uniform waveguide sections. More details about the fabrication process can be found in Supplementary Material \cite{seesupplementary}.

	\section{Acoustic design and characterization}

	\subsection{Acousto-optic interaction}\label{sec:AO theory}
Vibrations in optical waveguides interact with the propagating light through multiple processes. At the atomic scale, the \textit{photo-elastic} effect describes perturbations in the dielectric permittivity of a material as its crystal structure shrinks and expands under mechanical strain. Assuming without that the permittivity $\varepsilon$ is isotropic, it is modified by an amount \cite{chang1995acoustoopticdevices}
	\begin{equation}
		\delta\varepsilon_{ij} = -\varepsilon^2 p_{ijkl} S_{kl},
	\end{equation}
where $p_{ijkl}$ is the material's photo-elastic tensor and $S_{kl}$ is the strain tensor. Considering an optical mode $(E_i, H_i)$ propagating in a periodic structure along the y-axis and undergoing external strain, the variation of its propagation constant $\beta$ derived from first order perturbation theory is
	\begin{equation}\label{eq:PE}
	\delta\beta_\mathrm{PE} = \dfrac{\pi c\varepsilon_0}{\lambda_0}\dfrac{\iiint \varepsilon^2E_i^*E_j p_{ijkl}S_{kl}\mathrm d V}
	{\operatorname{Re}\iiint E_x^*H_y - E_y^*H_x \mathrm dV},
	\end{equation}
where the integrals are taken over a period of the structure, and $\lambda_0$ is the vacuum wavelength of light (see Appendix \ref{sec:perturbation_theory} for derivation). In addition to this microscopic effect, the guided optical mode is also affected by distortions of the waveguide geometry through the \textit{moving-boundary effect} contribution \cite{johnson2002perturbationtheory}, which accounts for displacement in air-dielectric and dielectric-dielectric interfaces:
	\begin{equation}\label{eq:MB}
		\delta\beta_\mathrm{MB} = \dfrac{\pi c\varepsilon_0}{\lambda_0}\dfrac{\iint (\mathbf u\cdot \mathbf n)(\Delta\varepsilon\lvert\mathbf E_\parallel\rvert^2 - \Delta(\varepsilon^{-1})\lvert \mathbf D_\perp\rvert^2)\mathrm d S}
		{\operatorname{Re}\iiint E_x^*H_y - E_y^*H_x \mathrm dV},
	\end{equation}
where the surface integral is taken over all material interfaces, $\mathbf u$ is the mechanical displacement field, $\mathbf n$ is the normal vector, and $\Delta\varepsilon$ $(\Delta(\varepsilon^{-1}))$ is the difference in (inverse) permittivity between the two materials. 
A weak second order electro-optic effect is also sometimes considered \cite{shao2019microwavetoopticalconversion}, but we neglect it here as both the piezoelectric and electro-optic coefficients of GaAs are small.	

The total phase shift over an interaction length $L_\mathrm{int}$ is then given by $\Delta\varphi = L_\mathrm{int}\delta\beta$ with $\delta\beta = \delta\beta_\mathrm{PE} + \delta\beta_\mathrm{MB}$. As the mechanical strain scales with the square root of the power carried by the acoustic wave per unit length, we choose to work with a normalized figure of merit
 \begin{equation}\label{eq:bpFOM}
 \tilde a_p = \frac{\delta\beta}{\sqrt{P_\mathrm{SAW}/L_\mathrm{int}}} = \frac{\Delta\varphi}{\sqrt{P_\mathrm{SAW}\,L_\mathrm{int}}}
 \end{equation}
 representing the acousto-optic modulation efficiency for unit acoustic power and interaction length in \si{rad\per \sqrt{\watt\,\meter}}. 
This normalized figure of merit is well suited to quantify the efficiency of the AO process itself regardless of the final device geometry and practical constraints, such as the IDT conversion efficiency. 
On the other hand, we can define a related figure of merit with closer links to experimental characterization of actual, fabricated devices that includes the final geometry and RF to acoustic power conversion $a_p$ \cite{xu2024unveilingefficient, delima2006compactmachzehnder, zhang2025onchip7, wan2022highlyefficient}, representing phase modulation depth for unit input electrical power. This quantity, expressed in \si{rad\per \sqrt{mW}}, is related to $\tilde a_p$ by:
  \begin{equation}\label{eq:apFOM}
 	a_p = \frac{\Delta\varphi}{\sqrt{P_\mathrm{RF}}} = \sqrt{\eta_\mathrm{IDT}L_\mathrm{int}}\tilde a_p
 \end{equation}
	where $P_\mathrm{RF}$ is the input electrical power and $\eta_\mathrm{IDT}=P_\mathrm{SAW}/P_\mathrm{RF}$ is the electrical to acoustic conversion efficiency of the IDT.

	\begin{figure}[t]
	\includegraphics{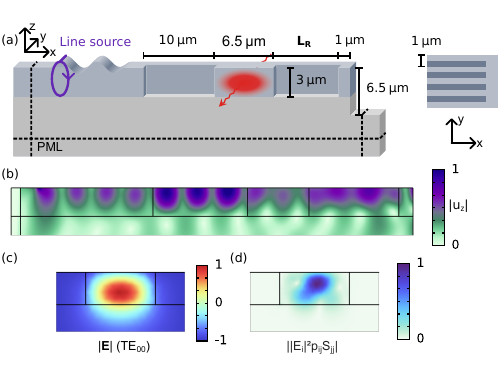}
	\caption{\label{fig:FEM_scheme} FEM simulation scheme. (a) Sketch of the acoustic FEM domain and geometrical parameters. (b) Simulated vertical displacement  amplitude $\lvert u_z\rvert$ near the surface for $L_\text{R}=\qty{9.5}{\micro m}$ at $\Omega/(2\pi)=\qty{459}{MHz}$. (c),(d) Simulated $\mathrm{TE_{00}}$ guided mode at vacuum wavelength $\lambda_0 = \qty{8.5}{\micro m }$ and corresponding photo-elastic overlap amplitude for the same parameters. All amplitudes are relative and taken on a y-cut plane of the simulated domain.}
\end{figure}
	
	\subsection{Simulation of acousto-optic coupling}\label{sec:design}

The acousto-optic devices are simulated using the finite-elements method (FEM, \texttt{COMSOL\textregistered}) in two separate electro-magnetic and acoustic 3D models, then combined through the overlap integrals of the previous section. Geometry is set to match experimental etching depths of \qty{6.5}{\micro m} in the trench regions and \qty{3}{\micro m} in the sub-$\lambda$ gratings. Materials parameters used in the simulations are given in appendix \ref{sec:material_properties}.
The electro-magnetic model is a standard mode analysis centered around the waveguide and spanning a full period of the sub-wavelength grating, and yielding the electromagnetic mode profile in the volume. 
The acoustic model simulates the displacement profile of the whole grating and waveguide structure. 
Acoustic excitation is performed using a \textit{prescribed displacement line} acting as a monochromatic acoustic source, reproducing the incident SAW motion (Fig.~\ref{fig:FEM_scheme}(a)). It is set at least $2\lambda_\text{SAW}$ away from the grating. 
Using this strategy allows one to decouple the acousto-optic design of the structure from the design of the IDTs. 
Since both the etched structure and the input SAW have an XZ-symmetry plane, we can limit the domain to a half-period (\qty{500}{nm}) along $\text Y$ with symmetry boundary conditions. The side and bottom domain boundaries are managed by Perfect Matching Layers (PML), tuned to absorb incident acoustic waves with minimal reflections \cite{khelif2010locallyresonant, gao2023onchiptightly}.
	
Figure \ref{fig:FEM_scheme}(b) displays a representative profile of the vertical displacement amplitude $\lvert u_z\rvert$ on a $\text Y$ cut-plane of the domain, showing that the SAW excited on the left propagates through the whole structure up to the optical waveguide. The clear beating pattern with periodicity equal to roughly half of the SAW wavelength $\lambda_\text{SAW}\approx \qty{6.3}{\micro m}$ indicates that strong reflections are taking place at the right edge, as well as at interfaces between bulk GaAs and sub-$\lambda$ grating interfaces. The maximal displacement amplitude is reached in the sub-$\lambda$ gratings, as they constitute the least stiff region of the structure.

To extract a quantitative information about efficiency of the AO interaction, one needs to normalize the permittivity change per unit of acoustic power. This in turns requires determining the power carried by the SAW. 
The line source generates both surface and bulk acoustic waves. The SAW profile is expected to be very close to that of the [110]-Rayleigh mode on bare (100)-GaAs, which takes the analytical form \cite{simon1996couplingsurface}
	\begin{equation}
		\left(\begin{array}{c}
			u_x \\ u_z
		\end{array}\right)(x,z,t) = 	\left(\begin{array}{c}
			2\operatorname{Re}\left(e^{i\Theta q z - i \varphi}\right) \\ 2i\operatorname{Im}\left(\gamma e^{i\Theta q z - i \varphi}\right)
		\end{array}\right)e^{iq(x-v_\text{SAW}t)}
	\end{equation}	
	with SAW phase velocity $v_\text{SAW}$, horizontal wave-vector $q=\Omega/v_\text{SAW}$ and complex constants $\Theta$, $\gamma$, $\varphi$ (see Appendix \ref{sec:suppl_normalization}). To minimize excitation of bulk waves, we thus prescribe a line source displacement polarity $(1, i\Gamma)$ with $\Gamma = \operatorname{Im}\left(\gamma e^{- i \varphi}\right)/\operatorname{Re}\left(e^{- i \varphi}\right) = -1.338$ to match that of the Rayleigh wave, such that only the desired SAW mode reaches the structure of interest.
In order to normalize the simulated quantities to the incident acoustic power as detailed in Section \ref{sec:AO theory}, we run a second instance of the simulation over a domain with identical meshing parameters in which the substrate is unetched, and record the surface displacement amplitude $\lvert u_z^{z=0}\rvert$ in the far field. The power carried by the SAW is then deduced from properties of the pure GaAs mode using Eq.~\ref{eq:suppl_SAW_power}.
	
Once the optical and acoustic profiles are obtained, the integrand of Eq.~\eqref{eq:PE} and \eqref{eq:MB} are computed (see e.g. Fig.~\ref{fig:FEM_scheme}(d) for the photo-elastic term), and integrated into a perturbation $\delta\beta$ of the optical mode propagation constant. 
This quantity is finally normalized by incident acoustic power into the modulation efficiency figure of merit $\tilde a_p$. Since the optical mode spatial distribution remains largely the same over the target wavelength range, the optical wavelength is fixed to $\lambda_0 = \qty{8.5}{\micro m}$ in the following paragraphs.

\begin{figure}
		\includegraphics{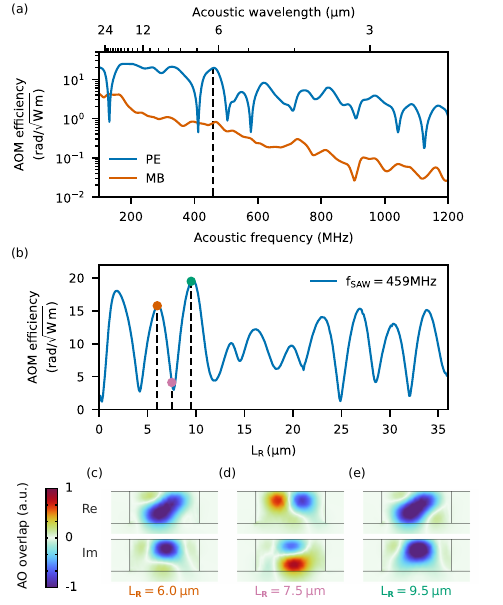}
		\caption{\label{fig:FEM_results} Simulation of the acousto-optic modulation (AOM) efficiency for $\lambda = 8.5$\si{\micro m} and TE polarization, normalized to unit interaction length and incident acoustic power. (a) Frequency response of the contributions from photo-elastic (PE) and moving-boundary (MB) effects for grating length $L_{\mathrm{R}}=\qty{9.5}{\micro m}$. The vertical dashed line at $\Omega/(2\pi)=\qty{459}{MHz}$ marks the frequency used in the rest of the simulations (b) Simulated efficiency as a function of $L_{\mathrm{R}}$. (c,d,e) Real and imaginary part of the photo-elastic overlap shown on a common arbitrary scale, at parameter values indicated by dashed lines in (b).}
	\end{figure}

The modulation efficiency for a structure with $L_{\mathrm{R}} = \qty{9.5}{\micro m}$ is plotted as a function of excitation frequency in Fig.~\ref{fig:FEM_results}(a) for the $\text{TE}_{00}$ guided optical mode, with contributions from the photo-elastic (PE, blue) and moving boundary (MB, orange) effects.
Since the latter is visibly weaker by at least an order of magnitude over the entire 100--\qty{1200}{MHz} simulated range, we can safely neglect it in what follows. 
	
Both efficiency curves show a broad downwards trend as the frequency increases and the SAW wavelength becomes smaller than the waveguide, causing destructive interference due to an uneven strain profile over the region occupied by the optical mode.
While the typical rule of thumb of setting the acoustic wavelength to at least twice the waveguide size to avoid this effect \cite{shao2019microwavetoopticalconversion, wang2022monolithicintegration} would suggest a maximum frequency of \qty{220}{MHz}, a broad peak in efficiency around $\Omega/(2\pi)=\qty{460}{MHz}$ is observed. We choose to target this region as a compromise between a high operating frequency and reasonable modulation efficiency.
	
	\subsection{Optimization of the acousto-optic coupling}

A closer examination of the simulated SAW profiles in Fig.~\ref{fig:FEM_scheme}(b) reveals that there is a strong acoustic reflection at the device edge, as well as weaker reflections at the boundaries between the subwavelength gratings and the un-etched GaAs regions. 
The waveguide thus forms a weak acoustic cavity, and interferences occur between the incoming and reflected SAW, explaining the broad resonance around 460~MHz. 
This implies that the overall device efficiency can be controlled by varying the reflection at the device edge, that is by changing the length $L_\text{R}$ of the right grating. We show in Fig~\ref{fig:FEM_results}(b) the results of the AOM efficiency simulations for a fixed SAW frequency of 459~MHz as $L_\text{R}$ is swept, showing a periodic dependency.  
The periodicity is approximately $\qty{3}{\micro m}$, matching a half-SAW wavelength. 
Figure \ref{fig:FEM_results}(c,d,e) show the full photo-elastic overlap profile, i.e.\ the integrand of Eq.~\ref{eq:PE}, at specific values of $L_\text{R}$. At the efficiency maxima $L_\text{R} = \qty{6.0}{\micro m}$ and \qty{9.5}{\micro m}, the profile shows mostly constructive interference over the whole optical mode area, while at the minimum $L_\text{R} = \qty{7.5}{\micro m}$ two opposite nodes are visible. Because the etch depth is only about $\lambda_\text{SAW}/2$, the observed behavior is not fully periodic due to scattered bulk waves in some geometries.


Based on the simulation results above, a series of 16 phase modulator devices with grating lengths varying from $L_\text{R}=\qty{6}{\micro m}$ to \qty{13.5}{\micro m} was fabricated. 
A minimum grating length of 6~\si{\micro m} is required to ensure a good optical mode confinement, as shorter grating lengths would lead to leakage into the outer region (see Supplementary Materials for insight on grating designs \cite{seesupplementary}). 
Each modulation region is associated with an IDT with periodicity $\Lambda = \qty{6.3}{\micro m}$, targeting an operating frequency of $\Omega_0 = \qty{459}{MHz}$. The IDT finger size is set a \qty{50}{\percent} filling factor, with $N_p=100$ pairs of electrodes and an effective aperture of $W=\qty{880}{\micro m}$ that sets the interaction length.
	
	\subsection{Vibrometry}
	
	\begin{figure*}[!ht]
		\includegraphics[]{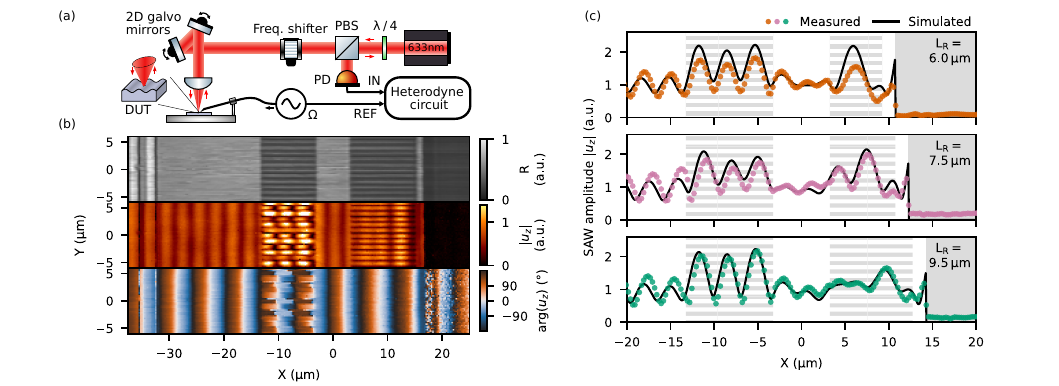}
		\caption{\label{fig:vibrometry} Acoustic characterization. (a) Vibrometry experimental setup used to measure acoustic activity induced on a device under test (DUT) driven at RF frequency $\Omega$. 
		(b) Vibrometry scan of the device with $L_{\mathrm{R}}=$\qty{12}{\micro m} driven at \qty{458}{MHz}. From top to bottom: measured local reflectivity, surface vertical displacement amplitude and phase. Vibrations are detected over the whole scanned region from the last two IDT electrodes on the left to the optical waveguide centered at $\mathrm{X=0}$ and the two gratings. (c) Measured surface vertical displacement amplitude $\lvert u_z\rvert$ from vibrometry scans averaged along the Y-axis (dots) and corresponding simulated profiles (lines), for devices with the same $L_{\mathrm{R}}$ values as in Fig.~\ref{fig:FEM_results}(c,d,e). Each curve is normalized so that $\lvert u_z\rvert$ averages to 1 in the $\mathrm{X<-\qty{15}{\micro m}}$ region.}
	\end{figure*}

	To evaluate the acoustic behavior of the devices independently from their piezoelectric and acousto-optic performance, the excited SAW profiles are probed in a home-made scanning heterodyne interferometry (or vibrometry) experiment (Fig.~\ref{fig:vibrometry}(a)). 
The vibrometer is based upon ref.~\cite{teyssieux2013absolutephase}, and relies on a folded heterodyne Michelson interferometer architecture, translating out-of-plane (vertical) displacement of the device surface into an optical signal captured by a fast photodiode. 

A linearly polarized 633~\si{\nano m} He:Ne laser beam is split in two using a $\lambda$/4 waveplate and a polarizing beam splitter, directing half of the beam on the photodetector to serve as the reference beam. The remaining half of the beam passes through a frequency shifter and a pair of galvo scanning mirrors, before being focused into a diffraction-limited spot on the sample. Reflection from the sample propagates back through the galvo mirrors and frequency shifter, and reflects back on the He:Ne laser output coupler to be directed towards the photodetector, interfering with the reference beam. 
When no SAW is excited on the sample, the resulting photodetected signal posses a strong beat at twice the frequency shifter drive frequency forming a carrier signal. SAW excitation at frequency $\Omega$ results in a small out-of-plane surface vibration, forming two sidebands shifted by $\pm \Omega$ from the carrier frequency. 
Feeding a portion of the electrical signal driving SAW excitation to a heterodyne I/Q demodulation circuit lets us retrieve the local amplitude and phase of vertical displacement $u_z$. More information on the setup can be found in the Supplementary Materials \cite{seesupplementary}. 

Figure \ref{fig:vibrometry}(b) shows a vibrometry scan for the $L_\text{R} = \qty{12}{\micro m}$ device with its IDT driven at $\Omega = \qty{458}{MHz}$. Top, middle and bottom panel respectively display the measured (relative) reflectivity, amplitude and phase of $u_z$ over a $62.5\times$\qty{12}{\micro m\squared} scanned surface with \qty{250}{nm} steps. 
SAW generated at the IDT (visible on the left) are observed propagating to the optical structure and reaching the optical waveguide at $\mathrm {X=\qty{0}{\micro m}}$. 
The standing wave pattern in displacement amplitude predicted in simulations is clearly visible here, with a periodicity matching half of the IDT period. 
Importantly, the acoustic phase at the waveguide is seen to be uniform along the Y-axis throughout the device length, ensuring a constructive effect of the acousto-optic interaction on transmitted mid-IR light. 

More information can be extracted by averaging the measured displacement amplitude over the Y-axis for equivalent scans of the three devices with grating lengths $L_\text{R}$ matching the ones of the simulations shown in Fig.~\ref{fig:FEM_results}(c,d,e). 
The results are shown in Figure \ref{fig:vibrometry}(c). 
The measured amplitude (dots) is overlaid on top of the simulated profile (solid lines). Each curve is normalized so that the average amplitude in the $X<15$~\si{\micro m} region is equal to 1. 
The agreement between experimental data and simulation is very good, and the general shape of the beating profile is well reproduced, with a moderate shift. 

We note however that while this validates the numerical model of the devices, even considering deviations from ideal geometry arising during fabrication, interpreting the vibrometry maps is not straightforward. 
Focusing on the waveguide section, one might expect the maxima of AOM efficiency for $L_\text{R} = \qty{6}{\um}$ and \qty{9.5}{\um} to simply correspond to a maximum of the displacement amplitude centered on the waveguide at $\mathrm{X=0}$ in Fig.~\ref{fig:vibrometry}(c). 
While simulations and vibrometry measurements show a similar surface profile, both agree that this is not the case. 
This is because the displacement standing wave pattern at a depth equal to half the core thickness $\mathrm Z = -\qty{1.5}{\micro m}$ (which is correlated with AOM efficiency) does not match that of the surface. 
We attribute this to scattered bulk waves related to the finite etch depth. 
Additionally, the lower beating contrast at the waveguide confirms the presence of weak acoustic cavity effects. 
		
As the lateral resolution of the vibrometry setup is limited by optical diffraction (\qty{0.57}{\micro m} FWHM diffraction limit, ${<\qty{1}{\micro m}}$ measured), measurements on the sub-$\lambda$ gratings with a width of \qty{500}{nm} cannot be considered fully quantitative, even though the expected beating profile from simulations is well reproduced. In particular an apparent increase in displacement amplitude can be observed if the focused laser beam hits both the top and etched surfaces of the sample \cite{shen2017phasesensitive,rembe2006laserscanningconfocal}.

Although care must be taken when interpreting the vibrometry maps, this characterization proves to be highly useful both to validate our simulation frame, and to ensure that no fabrication imperfection or defect lead to the scattering of the surface waves that would make the AOM efficiency vanish.

	\section{Mid-IR optical characterization}
	
	\subsection{Device and setup description}
	
	\begin{figure*}
	\includegraphics{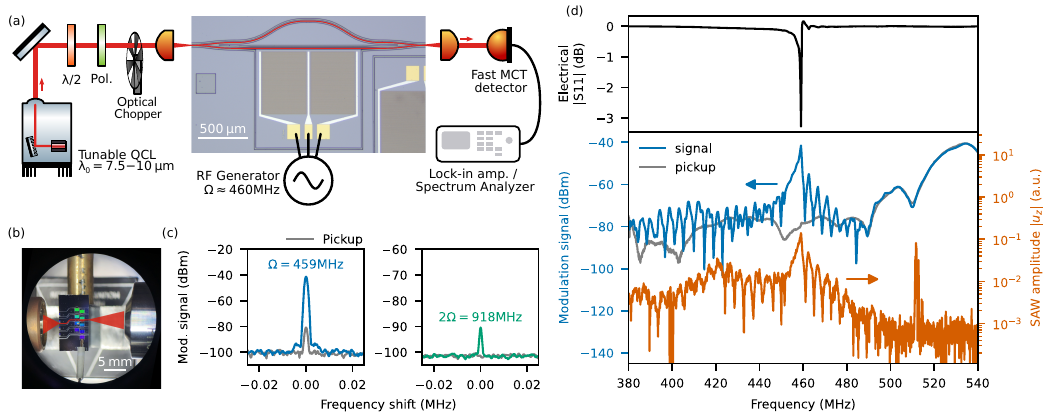}
	\caption{\label{fig:optical_setup} (a) Sketch of the setup used for optical characterization, including a microscopy image of a device with the waveguide highlighted in red. On-chip light is split then recombined in an unbalanced Mach-Zehnder interferometer, converting the phase modulation applied to a single arm into an amplitude modulation directly measurable by a fast photodetector. QCL: Quantum Cascade Laser. $ \lambda/2$, Pol.: half-wave plate and polarizer. (b) Picture of the free-space cleaved-faced coupling including a full sample and a pair of lenses, with the optical path sketched in red. A RF probe is visible at the bottom. (c) Photodetector signal at $\Omega$ and $2\Omega$ when driving the IDTs at $\Omega=\qty{459}{MHz}$, as captured by a spectrum analyzer. Laser wavelength is $\lambda_0 = \qty{8.0}{\micro m}$ and RF power at the generator is \qty{17}{dBm}. (d) Top: VNA measurement of the S11 reflectivity of an IDT, where a linear baseline has been subtracted. Bottom: Photodetected modulation signal level at $\lambda_0 = \qty{8.0}{\micro m}$ and \qty{14}{dBm} of injected RF power,  measured as a function of the drive frequency for the same device (left axis). Background level when the laser is off is shown in grey. Vertical surface displacement amplitude measured by vibrometry at the center of the waveguide (right axis), at \qty{20}{dBm} of injected power. Parasitic resonances from the vibrometry setup are visible at \qty{400}{MHz} and \qty{512}{MHz}.}
\end{figure*}

We now turn to the optoelectronic characterization of our integrated AOM devices. 
A full device is shown in the microscope image of Fig.~\ref{fig:optical_setup}(a). 
The acousto-optic phase modulation region is embedded into an unbalanced Mach-Zehnder interferometer (MZI), converting phase modulation into directly measurable amplitude modulation (see Supplementary Materials for a full sketch of the device cross-section \cite{seesupplementary}). 
The second arm is isolated from SAW activity by two fully etched grooves, as confirmed by vibrometry measurements. 
As will be detailed later, setting the length of the second arm to be longer by $\Delta L = \qty{90}{\micro m}$ than the modulated arm enables us to tune the interferometric phase difference by adjusting the optical wavelength. 
Coupling from free-space to the integrated devices is achieved using a 700~\si{\micro m} long linear taper with a 20~\si{\micro m} wide input. 
MZI couplers and transitions from ridge waveguide to sub-$\lambda$ grating section are designed as simple y-junctions and adiabatic transitions, and bending radii are maintained above \qty{500}{\micro m} to minimize radiative losses. 
	
The experimental setup for optical characterization is sketched in Fig.~\ref{fig:optical_setup}(b). 
The source is a continuous wave external cavity quantum cascade laser (Daylight MIRCAT), consisting of two laser chips spanning wavelengths from \qty{7.5}{\micro m} to \qty{10}{\micro m}. 
The naturally TM-polarized laser light is rotated to TE by a $\lambda/2$--polarizer pair then coupled into the device, and the transmitted light is directed to a fast \ce{HgCdTe} (MCT) amplified photodetector (Vigo PVI-4TE-10.6). 
Coupling to and from the sample is done through cleaved edges (Fig.~\ref{fig:optical_setup}(b)), using a pair of aspheric lenses (BD2 chalcogenide 6~mm and ZnSe 12~mm respectively). 
A signal generator (R\&S SMF100A) is used to drive the IDTs via an RF probe. 
The photodetected signal is fed to a spectrum analyzer for RF measurements (Keysight N9344C). 
For measurements of optical DC transmission through the sample, a lock-in amplifier (Stanford SR830) paired with an optical chopper placed before the sample replaces the spectrum analyzer.

	\subsection{Frequency response characterization.}
	
We perform an initial characterization of a device at a fixed wavelength of \qty{8.0}{\micro m}, driving the IDT at its design frequency $\Omega_0 = \qty{459}{MHz}$. 
Figure \ref{fig:optical_setup}(c) displays representative spectrum analyzer traces at $\Omega_0$ and $2\Omega_0$, evidencing light modulation at the IDT drive frequency and second harmonic. 
The background signal remaining when the light is blocked is attributed to pickup noise from the IDT acting as a radiating RF antenna. 
To gain further insight on the device, we measure its electrical frequency response around $\Omega_0$ and report it in Figs.~\ref{fig:optical_setup}(d). 
A first electrical reflectivity measurement of the $S_{11}$ parameter of the IDT conducted with a VNA is shown in the top panel with a simple linear baseline correction, displaying a sharp resonance at \qty{459}{MHz} corresponding to the expected Rayleigh SAW ($v_\text{SAW}=\Lambda \Omega_0= \qty{2.89}{km/s}$). 
The $\qty{-3}{dB}$ (50\%) magnitude constitutes a rough estimate of the maximum electrical to acoustic conversion efficiency $\eta_\mathrm{IDT}$ \cite{camara2017vectornetwork}. 
The bottom panel of Fig.~\ref{fig:optical_setup}(d) shows in blue the amplitude of the optical output at the modulation frequency $\Omega$ while sweeping the electrical drive frequency. 
The main features we observe are related to the IDT response, with a skewed sine cardinal shape typical for an IDT with non negligible reflectivity \cite{morgan2007surfaceacoustic}, and sidelobe periodicity $\qty{4.5}{MHz}\approx \Omega_0/{(2\pi N_p)}$ and FWHM \qty{0.6}{MHz}. 
Below 500~MHz, the photodetected signal drops below the pickup noise (RF drive on, laser blocked), which we interpret as destructive interference between the photodetected signal and pickup noise radiated by the IDT. 
We observe a huge increase in noise above 500~MHz masking all signal, the origin of which is not elucidated yet. 
Finally, a vibrometry measurement of the displacement amplitude frequency response in the middle of the waveguide shows a comparable shape, validating the acousto-optic nature of the observed optical modulation.

In order to asses quantitatively the performance of our AOM devices, we derive here the main equations governing signal detection in a MZI-AOM device. 
Assuming equal optical propagation losses in the two MZI arms, light transmitted through the device and incident to the photo-detector has optical intensity:
	\begin{equation}
	I(t) = \frac {I_0}2(1+\cos\Phi),
	\end{equation}
where $\Phi$ is the interferometric phase difference and $I_0$ is a reference intensity accounting for optical losses throughout the sample and experimental setup.
In the absence of electrical input power, $\Phi$ is simply the optical path difference between the two arms $\varphi_s = 2\pi n\Delta L/\lambda_0$. By tuning the laser wavelength, this static phase shift can be swept periodically. 
Adding a phase modulation with amplitude $\Delta\varphi$ at frequency $\Omega$, the total phase difference becomes:
	\begin{equation}
		\Phi(t) = \Delta\varphi\cos(\Omega t) + \varphi_s.
	\end{equation}
	
In the frequency domain, the light intensity is expanded using the Bessel functions of the first kind $J_n$ into harmonics of the modulation frequency \mbox{$I(t) = \sum_{n\geq 0} T_n \cos n\Omega t$}, with relative intensities
	\begin{align}
			T_0 &= I_0/2\left(1+\cos(\varphi_s) J_0(\Delta\varphi)\right)\\
	T_{n\geq 1} &= I_0\cos(\varphi_s + n\pi/2) J_n(\Delta\varphi).
	\end{align}	
	
	In particular at quadrature points where \mbox{$\varphi_s = \pi/2 + p\pi$} and in the small modulation depth limit where \mbox{$J_n(\Delta\varphi) \sim \Delta\varphi^n/(2^nn!)$}, $\Delta\varphi$ is obtained as  the ratio between signal at the modulation frequency and average intensity:
	\begin{equation}\label{eq:ratioT1T0}
		\Delta\varphi = \left\lvert\frac{T_1}{T_0}\right\rvert_{\mid \varphi_s = \pi/2+p\pi}
	\end{equation}

	\begin{figure}
		\includegraphics{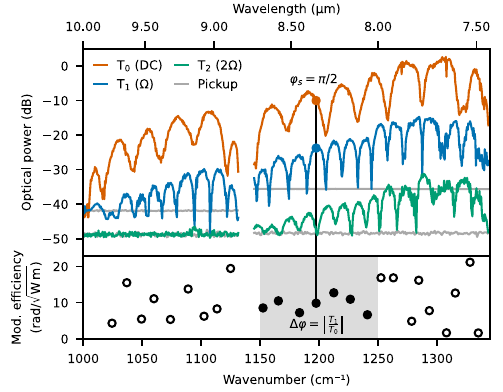}
		\caption{\label{fig:wavelength_sweep} Transmission spectra of the $L_{\mathrm{R}} = \qty{7}{\micro m}$ device for an RF drive of $\qty{14}{dBm}$ at $\Omega/(2\pi)=\qty{459}{MHz}$. The three oscillating curves show from top to bottom: DC transmission (optical chopper measurement), and signals measured at $\Omega$ and $2\Omega$ using a spectrum analyzer. 
Gray curves show the background level for RF measurements with RF drive switched on and laser off. 
Bottom: Calculated modulation efficiency at each maximum of the first harmonic obtained from the ratio of harmonic 1 to the DC component, normalized by injected RF power and IDT electro-acoustic conversion loss. Filled circles denote the wavelength range retained for further analysis.}
	\end{figure}

	We thus show in Fig.~\ref{fig:wavelength_sweep} the measured $T_{0,1,2}$ as a function of the laser wavenumber when driving the device at the peak IDT frequency, with $\qty{14}{dBm}$ of input RF power. The grey curves show background noise level with laser blocked, but RF drive on. Each curve is normalized accounting for the experimental frequency-dependent responsivity of the amplified photodetector (see Supplementary Materials \cite{seesupplementary}), and the DC transmission value is obtained from the chopped measurement accounting for the square signal conversion factor $\sqrt{2/\pi}$. 
The three curves are in good agreement with the expected behavior of $T_{0,1,2}$ for pure phase modulation across the whole 7.5--$\qty{10}{\micro m}$ wavelength range. The general envelope of the signal is dominated by the gain curve of each laser chip, with an additional trend of decreased transmission at larger wavelength due to passive waveguide losses. A small gap around 1140~cm$^{-1}$ separates the gain domains of the two QCL chips, with decreased signal at the edges of each region corresponding to operation close to threshold.  

From this, the phase modulation depth is calculated at each maximum of $T_1$ using Eq.~\eqref{eq:ratioT1T0}. 
Converting the modulation depth to an acousto-optic modulation efficiency $\tilde a_p$ expressed in the same unit as the simulated ones requires normalization by the interaction length, that is the IDT aperture of $\qty{880}{\micro m}$, and by the acoustic power (Eq.~\ref{eq:bpFOM}).
The latter is estimated from the output power of the RF generator ($\qty{14}{dBm}$), accounting for \qty{1}{dB} losses in RF cables and microprobe, then dividing by the electro-acoustic conversion rate $\eta_\mathrm{IDT}\approx1-\lvert S_{11}\left(\Omega_0\right)\rvert^2$ for each individual IDT associated with each device (between $0.35$ and $0.6$ with an average of $0.5$). 
The resulting AOM efficiency in \si{rad/\sqrt{W\,m}} is plotted in the lower panel of Fig~\ref{fig:wavelength_sweep}. 
It remains relatively stable on average throughout the wavelength range, though greater variations are observed below $\qty{8}{\micro m}$ where the laser exhibits less stable behaviour (we attribute this to stronger retro-injection into the laser chip), and for the second chip covering the range above $\qty{8.75}{\micro m}$ as the passive waveguide causes greater losses for longer wavelengths.

\begin{figure}
		\includegraphics{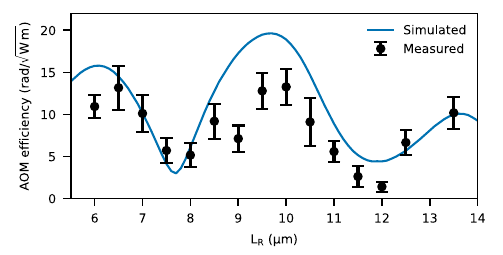}
		\caption{\label{fig:aom_efficiency} Modulation efficiency of the 16 measured devices averaged over the 1150--\qty{1250}{cm^{-1}} range each at frequency of peak efficiency of their IDT ($\num[separate-uncertainty = true]{459.0(2)}\,\si{MHz}$), compared to FEM simulations (solid line, same as \autoref{fig:FEM_results}(b)). Error bars correspond to the standard deviation over the wavelength range and multiple measurement runs.}
	\end{figure}

We repeat this experiment for all 16 fabricated devices, measuring the full $T_{0,1,2}$ as a function of the laser wavenumber when driving the device at the peak IDT frequency and extracting the modulation efficiency at each peak of $T_1$ (additional data is provided in Supplementary Material \cite{seesupplementary})
To study the effect of $L_{\mathrm{R}}$, we then report in Fig.~\ref{fig:aom_efficiency} the average of the modulation efficiency for each device over the central 8--\qty{8.75}{\micro m} range (shaded area in Fig.~\ref{fig:wavelength_sweep}, points shown in full circles), repeating the experiment 2 to 3 times depending on the device.
Data points thus represent the average modulation efficiency, and the errorbars width shows the standard deviation of the series. 
The measurements are compared to the simulated AOM efficiency extracted from Fig.~\ref{fig:FEM_results} and restricted to the appropriate $L_{\mathrm{R}}$ range (solid blue line), without any correction factor. 
We observe a quantitative agreement between the simulated and measured efficiencies, validating our approach. 
The expected efficiency oscillation as a function of grating length $L_\text{R}$ is clearly present in the experimental data, with the best device at $L_\text{R}=\qty{10}{\um}$ reaching a normalized efficiency of $\tilde a_p= \qty{9.4}{mrad\per\sqrt{W\,m}}$.

Taking into account the interaction length of \qty{880}{\micro \meter} and the IDT efficiency $\eta_\mathrm{IDT} =\qty{46}{\percent}$, the corresponding standard figure of merit for device performance in modulation depth per unit input RF power (Eq.~\ref{eq:apFOM}) is $a_p = \qty{6.0}{mrad/\sqrt{mW}}$, i.e. \qty{3.4}{\degree} at \qty{20}{dBm}. 
These values are consistent with previous reports on GaAs/AlGaAs integrated AOM \cite{delima2006compactmachzehnder}, who reported a device with $\qty{73}{mrad/\sqrt{mW}}$ at $\lambda_0 = \qty{950}{nm}$: considering that the photo-elastic interaction scales with $1/\lambda_0$ (Eq.~\ref{eq:PE}), this factor 10 difference in efficiency can be mostly attributed to the ratio of optical wavelengths.
For reference, state of the art devices at telecom wavelengths, based on Lithium Niobate platforms with thoroughly optimized geometries as well as near-perfect piezoelectric matching can reach around $\qty{200}{mrad/\sqrt{mW}}$ per MZI arm \cite{wan2022highlyefficient}.

	\section{Conclusions}
In this work, we established a simulation and design framework for integrated acousto-optic phase modulation in the 7.5--\qty{10}{\micro m} range at \qty{460}{MHz}, operating in a regime where SAW and optical wavelength are comparable. 
The similar dimensions of acoustic and optical wavelengths leads to interesting challenges of co-confinement of acoustic and electromagnetic fields, whether aiming for high frequencies or like here large optical wavelengths. 
An important aspect of this work is the quantitative acousto-optic efficiency simulation, in which a proper normalization of the acoustic power is enabled by additional strain profiles calculated in reference geometries without microstructures, combined to the analytical derivation of acoustic power carried by the SAW on GaAs.
Our quantitative simulation scheme proved to accurately predict complex acoustic behaviors in the proposed devices, showing remarkable agreement with both vibrometric acoustic characterization and opto-electronic device performance. 
Inserting the AOM in an unbalanced MZI allowed us to properly determine the main efficiency figure of merit, with the best device showing an AOM efficiency of $\qty{9.4}{mrad\per\sqrt{W\,m}}$. 
While the measured performances are not on par with comparable devices in the near-IR, these devices constitute a first step in exploiting interactions between acoustic waves and light at the far end of the infrared spectrum. 
There are indeed many valid strategies to improve this result, depending on application, that can also be combined.
As we targeted the demonstration of pure phase modulation here, we chose not to integrate the AOM directly in an \textit{optical} microcavity, which would produce amplitude modulation, but this has been shown to drastically enhance the AOM conversion efficiency \cite{shao2019microwavetoopticalconversion, wan2025hybridthinfilm}. Another route would be to work on a good \textit{acoustic} cavity around the waveguide (using e.g. a phononic crystal structure \cite{laude2015phononiccrystals}) to increase the acoustic power at the waveguide \cite{hassanien2021efficientwideband, zhou2024electricallyinterfaced}. 
Finally, heterogenous integration of strong piezoelectric materials is also possible, under the additional constraint of isolating them from the optical waveguide to avoid any optical losses caused by these materials in the mid-IR.

		\begin{acknowledgments}
This work was partially supported by the French National Research Agency: project BIRD (Grant No. ANR-21-CE24-0013), Project Acousto-MIR (No. ANR-24-CE24-1691), PIA funding ANR-20-IDEES-0002, national France 2030 funding (PhOM – Graduate School Physique and ANR-11-IDEX-0003), and by the European Union through FET-Open Grant NEHO (Grant No. 101046329), and ERC under grant SMART-QDEV (Grant No. 101095959).
This work was done within the C2N micro nanotechnologies platforms and partly supported by the RENATECH network and the General Council of Essonne. 
We acknowledge the technical help of the C2N cleanroom staff, and in particular Alan Durnez and Laetitia Leroy for their assistance in PECVD and metal deposition. We thank Jean-Michel Manceau and Carlos Ramos for continuous fruitful scientific discussions.  
Automation of experiments was made possible using \texttt{Autolab} \cite{2026autolabprojectautolab} and \texttt{GnuRadio} \cite{gnuradio, blossom2004gnuradio}.

		\end{acknowledgments}

		\section*{Authors contributions}
		M.\ Morassi and A.\ Lemaître performed the epitaxy of the heterostructure. T.\ Gérodou and M.\ Jeannin fabricated the samples, with the help of J.-R.\ Coudevylle, E.\ Herth and A. Bousseksou. 
		T. Gérodou and M. Jeannin built the heterodyne interferometer, assisted by D. Teyssieux. 
		L. Lucia, M. Urbain and A. Bousskesou built the integrated characterization setup. 
		T. Gérodou performed all experimental characterizations, with the participation of M.\ Urbain and L. Lucia for integrated device characterization. 
		T.\ Gérodou analyzed the raw data from all experiments, and performed the numerical simulations, under the supervision of M. Jeannin. 
		R. Colombelli, A. Bousseksou and M. Jeannin acquired fundings. 
		M.\ Jeannin conceived and directed the project. All authors discussed the results and commented the manuscript. 
		
		\section*{Data availability}
Data and code supporting the findings of this study are publicly available \cite{gerodou2026supportingdata}.

		\appendix
				
		\section{Derivation of acousto-optic couplings.}\label{sec:perturbation_theory}
		
		The standard perturbation theory for an optical mode $(\mathbf E, \mathbf H)$ at frequency $\omega$ under a small variation $\delta\varepsilon$ of dielectric permittivity (due to the photo-elastic effect) is\cite{snyder1984opticalwaveguide}:
		\begin{equation}
			\delta\omega = -\frac{\omega}{2}\frac{\langle \mathbf E \rvert \delta\varepsilon \lvert \mathbf E \rangle}{\langle \mathbf E \rvert \varepsilon \lvert \mathbf E \rangle}
		\end{equation}
		For a permittivity with smooth spatial distribution, the numerator is simply the volume integral $\iiint E^*_{i} \delta\varepsilon_{ij} E_j \mathrm d V$. However, the case of an optical waveguide under mechanical strain is not fully captured by this term, as small shifts in sharp air-dielectric and dielectric-dielectric interfaces also induce a first order perturbation of the optical mode \cite{johnson2002perturbationtheory}. The theory is restored by an additional surface integral $\iint (\mathbf u\cdot \mathbf n)(\Delta\varepsilon\lvert\mathbf E_\parallel\rvert^2 - \Delta(\varepsilon^{-1})\lvert \mathbf D_\perp\rvert^2)\mathrm d S$ taken over material interfaces, where $\mathbf u$ is the mechanical displacement field, $\mathbf n$ is the normal vector directed from material 1 to 2, and $\Delta\varepsilon$ $(\Delta(\varepsilon^{-1}))$ is the difference in (inverse) permittivity between 2 and 1.
		
		The two contributions Eq.~\ref{eq:PE} and \ref{eq:MB} in the main text are expressed in terms of variation of the optical wavevector using the following expression of group velocity in a non-dispersive dielectric \cite[ch. 11]{snyder1984opticalwaveguide}:
		\begin{equation}
			\frac{\delta\omega}{\delta\beta} = \frac{1}{\varepsilon_0}\frac{\iiint (\mathbf E^* \times \mathbf H)\cdot \hat{\mathbf y}\mathrm d V}{\iiint  \varepsilon \lvert \mathbf E \rvert^2 d V}
		\end{equation}
		as $\delta\beta = \delta\omega/({\delta\omega}/{\delta\beta})$.
		
		\section{Material properties}\label{sec:material_properties}
		We used the following material properties for $\ce{GaAs/Al_xGa_{1-x}As}$ heterostructures in FEM simulations (Voigt notation):
		\begin{table}[h!]
		\begin{ruledtabular}
	\begin{tabular}{ccc}
		Density \cite{goldberg1996aluminiumgallium} (\si{\gram/\cm\cubed}) &  & $5.32-1.56x$ \\
		Relative permittivity \cite{goldberg1996aluminiumgallium}&  & $10.89-2.73x$ \\
		Elastic stiffness (GPa) \cite{adachi1985gaasalas}& $C_{11}$ & $118.8 - 1.4x$ \\
		& $C_{12}$ & $53.8 - 3.2x$ \\
		& $C_{44}$ & $59.4 - 0.5x$ \\
		Photo-elastic tensor \cite{chang1995acoustoopticdevices}& $p_{11}$ & $-0.165$ \\
		& $p_{12}$ & $-0.14$ \\
		& $p_{44}$ & $-0.072$ \\
	\end{tabular}
\end{ruledtabular}
		\end{table}
		
		The $\ce{AlGaAs}$ layer was set to $x=0.5$. Photo-elastic parameters correspond to experimental measurements on pure GaAs at $\lambda = \qty{1.15}{\micro m}$, although these values could be wavelength-dependent. All the above parameters are given in the cubic crystal frame. The overlap integrals, which are performed in the [110] device coordinate system (c.f.\ Fig.~\ref{fig:FEM_scheme}(a)), require photo-elastic tensor elements rotated by $\pi/4$:
		\begin{equation}
			[p']_{xyz} = \left(\begin{array}{cccccc}
				p_{11}' & p_{12}' & p_{12} &  &  &  \\
				p_{12}' & p_{11}' & p_{12} &  &  &  \\
				p_{12} & p_{12} & p_{11} &  &  &  \\
				&  &  & p_{44}'&  &  \\
				&  &  &  & p_{44}' &  \\
				&  &  &  &  & p_{44}
			\end{array}\right)
		\end{equation}
		with $p_{11}' =  \frac12(p_{11} + p_{12} + 2 p_{44}) = -0.2245$, $p_{12}' =  \frac12(p_{11} + p_{12} - 2 p_{44}) = -0.0805$ and $p_{44}' =  \frac12(p_{11} - p_{12}) = -0.0125$.
		\section{Normalization in FEM simulations}\label{sec:suppl_normalization}
		Following \cite{simon1996couplingsurface}, the [110]-Rayleigh SAW on bare (100)-GaAs with wave-vector $q$ takes the form
		\begin{equation}\label{eq:rayleigh_displ}
			\left(\begin{array}{c}
				u_x \\ u_z
			\end{array}\right)(x,z,t) = 	\left(\begin{array}{c}
			2A\operatorname{Re}\left(e^{i\Theta q z - i \varphi}\right) \\ 2iA\operatorname{Im}\left(\gamma e^{i\Theta q z - i \varphi}\right)
			\end{array}\right)e^{iq(x-v_\text{SAW}t)}
		\end{equation}	
		with $v_\text{SAW} = \Omega/q=\qty{2.86}{km/s}$ the SAW phase velocity, $\Theta = 0.48-0.5j$, $\gamma = -1.15-0.68$ and $\varphi = \qty{-2.09}{rad}$ (using material parameters from Appendix \ref{sec:material_properties}).
		
		To obtain the acoustic power carried by this SAW, we turn to the elastic Poynting vector $P_j = -\frac12(\partial_t u_i)c_{ijkl}S^*_{kl}$ as defined in \cite{auld1973acousticfields}. In the $xyz$ device frame, the $x$ component is 
		\begin{equation}
			P_x = \frac12iqv_\text{SAW}(u_1(C_{11}'S_{11}^*+C_{12}S_{33}^*)+2u_3C_{44}'S_{13}^*)
		\end{equation}
		where $C_{11}' = \frac12(C_{11} + C_{12} + 2 C_{44})$ and $C_{44}' =  \frac12(C_{11} - C_{12})$. Substituting Eq.~\ref{eq:rayleigh_displ} yields after some algebra
		\begin{align}
	P_x = q^2\lvert A\rvert^2v&\operatorname{Re}\left[\left(C_1'+C_2\gamma\Theta-(C_2+C_3)\gamma(\gamma+\Theta)\right)\Phi^2\right] \nonumber \\ + q^2\lvert A\rvert^2v&\operatorname{Re}\left[C_1'+C_2\gamma\Theta + (C_2+C_3)\gamma^*(\gamma+\Theta)\right]\lvert \Phi\rvert^2
	\end{align}
		where $\Phi = e^{iq\Theta z - i\varphi}$. Using $\int_{-\infty}^0\textup dz\Phi^2=e^{-2i\varphi}/(2iq\Theta)$ and $\int_{-\infty}^0\textup dz\lvert\Phi\rvert^2=-1/(2q\operatorname{Im}\Theta)$, the total acoustic power carried by the Rayleigh SAW per wavefront width is
		\begin{equation}
			P_l = \int_{-\infty}^0 P_x \mathrm d z = \lvert A\rvert^2 \Omega \tilde C 
		\end{equation}
		with 
				\begin{equation}\label{eq:ctilde}
					\begin{split}\tilde C = \frac12\operatorname{Re}&\left[\frac{(C_1'+C_2\gamma\Theta-(C_2+C_3)\gamma(\gamma+\Theta))e^{-2i\varphi}}{i\Theta}\right. \\
						-&\left. \frac{C_1'+C_2\gamma\Theta+(C_2+C_3)\gamma^*(\gamma+\Theta)}{\operatorname{Im}\Theta}\right].
					\end{split}
				\end{equation}
		For [110]-GaAs, $\tilde C = \qty{169}{GPa}$. Combining Eq.~\ref{eq:rayleigh_displ} and \ref{eq:ctilde}, we finally get the relation between linear wave power and vertical displacement amplitude at $z=0$:
		\begin{equation}\label{eq:suppl_SAW_power}
			P_l = \lvert u_z^{z=0}\rvert^2 \tilde C_z \Omega
		\end{equation}
		with $\tilde C_z = \tilde C / (4\lvert \operatorname{Im}\gamma (e^{-i\varphi})\rvert^2) = \qty{95.6}{GPa}$ in our case. For example, a $\qty{500}{MHz}$ wave with $\lvert u_z^{z=0}\rvert =\qty{1}{\text{\AA}}$ surface vertical amplitude carries approximately $\qty{3}{mW}$ of power per mm of wavefront.

	\bibliography{bibliography}

@misc{2026autolabprojectautolab,
	title = {Autolab-Project/Autolab},
	year = 2026,
	month = aug,
	publisher = {Autolab},
	url = {https://github.com/autolab-project/autolab},
	urldate = {2026-10-07},
	copyright = {GPL-3.0}
}

@article{adachi1985gaasalas,
	title = {{{GaAs}}, {{AlAs}}, and {{AlxGa1}}-{{xAs}}: {{Material}} Parameters for Use in Research and Device Applications},
	shorttitle = {{{GaAs}}, {{AlAs}}, and {{AlxGa1}}-{{xAs}}},
	author = {Adachi, Sadao},
	year = 1985,
	month = aug,
	journal = {Journal of Applied Physics},
	volume = {58},
	number = {3},
	pages = {R1-R29},
	issn = {0021-8979},
	doi = {10.1063/1.336070},
	urldate = {2026-09-15}
}

@incollection{auld1973acousticfields,
	title = {Acoustic Fields and Waves in Solids. 1},
	author = {Auld, Bertram A.},
	year = 1973,
	publisher = {Wiley},
	address = {New York},
	isbn = {978-0-471-03700-2},
	langid = {english}
}

@misc{blossom2004gnuradio,
	title = {{{GNU}} Radio: Tools for Exploring the Radio Frequency Spectrum},
	shorttitle = {{{GNU}} Radio},
	author = {Blossom, Eric},
	year = 2004,
	month = jun,
	url = {https://www.linuxjournal.com/article/7319}
}

@article{bonazzi2024metamaterialunipolar,
	title = {Metamaterial Unipolar Quantum Optoelectronics for Mid-Infrared Free-Space Optics},
	author = {Bonazzi, T. and Dely, H. and Didier, P. and Gacemi, D. and Fix, B. and Beck, M. and Faist, J. and Harouri, A. and Sagnes, I. and Grillot, F. and Vasanelli, A. and Sirtori, C.},
	year = 2024,
	month = nov,
	journal = {APL Photonics},
	volume = {9},
	number = {11},
	pages = {110801},
	issn = {2378-0967},
	doi = {10.1063/5.0225920},
	urldate = {2026-09-16}
}

@article{bouchoule2008anisotropicsmooth,
	title = {Anisotropic and {{Smooth Inductively Coupled Plasma Etching}} of {{III-V Laser Waveguides Using HBr-O}}[Sub 2] {{Chemistry}}},
	author = {Bouchoule, S. and Azouigui, S. and Guilet, S. and Patriarche, G. and Largeau, L. and Martinez, A. and Le Gratiet, L. and Lemaitre, A. and Lelarge, F.},
	year = 2008,
	journal = {Journal of The Electrochemical Society},
	volume = {155},
	number = {10},
	pages = {H778},
	issn = {00134651},
	doi = {10.1149/1.2965790},
	urldate = {2026-01-30},
	langid = {english}
}

@article{bricout2026pulsegeneration,
	title = {Pulse {{Generation}} by {{On-Chip Dispersion Compensation}} at 8 \textmu m {{Wavelength}}},
	author = {Bricout, Annabelle and Bertrand, Mathieu and T{\"a}schler, Philipp and Schneider, Barbara and Turpaud, Victor and Calcaterra, Stefano and Impelluso, Davide and Faverzani, Marco and Bouville, David and Coudevylle, Jean-Ren{\'e} and Edmond, Samson and Herth, Etienne and {Alonso-Ramos}, Carlos and Vivien, Laurent and Frigerio, Jacopo and Isella, Giovanni and Faist, J{\'e}r{\^o}me and {Marris-Morini}, Delphine},
	year = 2026,
	journal = {Laser \& Photonics Reviews},
	volume = {20},
	number = {8},
	pages = {e01485},
	issn = {1863-8899},
	doi = {10.1002/lpor.202501485},
	urldate = {2026-09-16},
	copyright = {\copyright{} 2025 The Author(s). Laser \& Photonics Reviews published by Wiley-VCH GmbH},
	langid = {english}
}

@article{camara2017vectornetwork,
	title = {Vector Network Analyzer Measurement of the Amplitude of an Electrically Excited Surface Acoustic Wave and Validation by {{X}}-Ray Diffraction},
	author = {Camara, I. S. and Croset, B. and Largeau, L. and Rovillain, P. and Thevenard, L. and Duquesne, J.-Y.},
	year = 2017,
	month = jan,
	journal = {Journal of Applied Physics},
	volume = {121},
	number = {4},
	pages = {044503},
	issn = {0021-8979, 1089-7550},
	doi = {10.1063/1.4974947},
	urldate = {2024-09-13},
	langid = {english}
}

@incollection{chang1995acoustoopticdevices,
	title = {Acousto-Optic Devices and Applications},
	booktitle = {Handbook of Optics},
	author = {Chang, I. C.},
	year = 1995,
	volume = {2},
	pages = {12.1-12.54},
	publisher = {McGraw-Hill New York}
}

@article{delima2006compactmachzehnder,
	title = {Compact {{Mach-Zehnder}} Acousto-Optic Modulator},
	author = {De Lima, M. M. and Beck, M. and Hey, R. and Santos, P. V.},
	year = 2006,
	month = sep,
	journal = {Applied Physics Letters},
	volume = {89},
	number = {12},
	pages = {121104},
	issn = {0003-6951, 1077-3118},
	doi = {10.1063/1.2354411},
	urldate = {2024-02-22},
	langid = {english}
}

@article{dely202210gbit,
	title = {10 {{Gbit}} S{\textsuperscript{-1}} {{Free Space Data Transmission}} at 9 \textmu m {{Wavelength With Unipolar Quantum Optoelectronics}}},
	author = {Dely, Hamza and Bonazzi, Thomas and Spitz, Olivier and Rodriguez, Etienne and Gacemi, Djamal and Todorov, Yanko and Pantzas, Konstantinos and Beaudoin, Gr{\'e}goire and Sagnes, Isabelle and Li, Lianhe and Davies, Alexander Giles and Linfield, Edmund H. and Grillot, Fr{\'e}d{\'e}ric and Vasanelli, Angela and Sirtori, Carlo},
	year = 2022,
	month = feb,
	journal = {Laser \& Photonics Reviews},
	volume = {16},
	number = {2},
	pages = {2100414},
	issn = {1863-8880, 1863-8899},
	doi = {10.1002/lpor.202100414},
	urldate = {2026-09-16},
	langid = {english}
}

@article{didier2026thinfilm,
	title = {Thin Film Lithium Niobate on Sapphire for Integrated Mid-Infrared Modulator},
	author = {Didier, Pierre and Jain, Prakhar and Bertrand, Mathieu and Kellner, Jost and Pitz, Oliver and Dai, Zhecheng and Kuttner, Tristan and Beck, Mattias and Chen, Baile and Faist, J{\'e}r{\^o}me and Grange, Rachel},
	year = 2026,
	month = feb,
	journal = {Nature Communications},
	volume = {17},
	number = {1},
	pages = {3050},
	publisher = {Nature Publishing Group},
	issn = {2041-1723},
	doi = {10.1038/s41467-026-69880-5},
	urldate = {2026-09-09},
	copyright = {2026 The Author(s)},
	langid = {english}
}

@article{gao2023onchiptightly,
	title = {On-{{Chip Tightly Confined Guiding}} and {{Splitting}} of {{Surface Acoustic Waves Using Line Defects}} in {{Phononic Crystals}}},
	author = {Gao, Feng and Benchabane, Sarah and Bermak, Amine and Dong, Shurong and Khelif, Abdelkrim},
	year = 2023,
	journal = {Advanced Functional Materials},
	volume = {33},
	number = {14},
	pages = {2213625},
	issn = {1616-3028},
	doi = {10.1002/adfm.202213625},
	urldate = {2026-10-24},
	copyright = {\copyright{} 2023 Wiley-VCH GmbH},
	langid = {english}
}

@misc{gerodou2026supportingdata,
	title = {Supporting Data for: {{Mid-Infrared Integrated Acousto-Optic Phase Modulation}} on {{GaAs}}/{{AlGaAs}}},
	shorttitle = {Supporting Data For},
	author = {G{\'e}rodou, Thomas and Urbain, Mathilde and Lucia, Luca and Herth, Etienne and Coudevylle, Jean-Ren{\'e} and Morassi, Martina and Lema{\^i}tre, Aristide and Teyssieux, Damien and Colombelli, Raffaele and Bousseksou, Adel and Jeannin, Mathieu},
	year = 2026,
	month = nov,
	publisher = {Zenodo},
	doi = {10.5281/zenodo.24334544},
	urldate = {2026-11-18}
}

@misc{gnuradio,
	title = {{{GNU Radio}}: {{A}} Free \& Open-Source Software Development Toolkit for Software-Defined Radios},
	year = 2001,
	url = {https://www.gnuradio.org}
}

@incollection{goldberg1996aluminiumgallium,
	title = {Aluminium {{Gallium Arsenide}}},
	shorttitle = {{{ALUMINIUM GALLIUM ARSENIDE}} ({{AlxGa1}}?},
	booktitle = {Handbook {{Series}} on {{Semiconductor Parameters}}},
	author = {Goldberg, {\relax Yu}. A.},
	year = 1996,
	month = nov,
	pages = {1--36},
	publisher = {World Scientific},
	doi = {10.1142/9789812832085_0001},
	urldate = {2026-09-15},
	isbn = {978-981-02-2935-1}
}

@article{hakl2021ultrafastquantumwell,
	title = {Ultrafast {{Quantum-Well Photodetectors Operating}} at 10 {$\mu$}m with a {{Flat Frequency Response}} up to 70 {{GHz}} at {{Room Temperature}}},
	author = {Hakl, Michael and Lin, Quyang and Lepillet, Sylvie and Billet, Maximilien and Lampin, Jean-Fran{\c c}ois and Pirotta, Stefano and Colombelli, Raffaele and Wan, Wenjian and Cao, J. C. and Li, Hua and Peytavit, Emilien and Barbieri, Stefano},
	year = 2021,
	month = jan,
	journal = {ACS Photonics},
	volume = {8},
	number = {2},
	pages = {464--471},
	issn = {2330-4022},
	doi = {10.1021/acsphotonics.0c01299},
	urldate = {2026-08-20}
}

@article{hassanien2021efficientwideband,
	title = {Efficient and Wideband Acousto-Optic Modulation on Thin-Film Lithium Niobate for Microwave-to-Photonic Conversion},
	author = {Hassanien, Ahmed E. and Link, Steffen and Yang, Yansong and Chow, Edmond and Goddard, Lynford L. and Gong, Songbin},
	year = 2021,
	month = jul,
	journal = {Photonics Research},
	volume = {9},
	number = {7},
	pages = {1182},
	issn = {2327-9125},
	doi = {10.1364/PRJ.421612},
	urldate = {2026-02-21},
	langid = {english}
}

@article{huertas-pedroche2025highspeedintegrated,
	title = {High-Speed Integrated Silicon Germanium Optical Modulator with 3 {{GHz}} Bandwidth in the 5-9 \textmu m Wavelength Range},
	author = {{Huertas-Pedroche}, Javier and Turpaud, Victor and Lucia, Luca and Bousseksou, Adel and Nguyen, Thi Hao Nhi and Dely, Hamza and Ngo, Gia Long and Calcaterra, Stefano and Impelluso, Davide and De Cerdeira Oliveira, Afonso and Herth, Etienne and Bouville, David and Edmond, Samson and Coudevylle, Jean Rene and Merupo, Victor and Peytavit, Emilien and Lampin, Jean-Fran{\c c}ois and Barbieri, Stefano and Frigerio, Jacopo and Colombelli, Raffaele and Isella, Giovanni and {Marris-Morini}, Delphine},
	year = 2025,
	month = oct,
	journal = {Optics Express},
	volume = {33},
	number = {20},
	pages = {42408},
	issn = {1094-4087},
	doi = {10.1364/OE.572563},
	urldate = {2026-02-15},
	langid = {english}
}

@article{huertas-pedroche2026integratedphase,
	title = {Integrated Phase Shifter Operating at 8\,{$\mu$}m Wavelength with 4\,{{kHz}} Bandwidth},
	author = {{Huertas-Pedroche}, Javier and Turpaud, Victor and Pinto, Tiago and Bricout, Annabelle and {El-Bouchikhi}, Nooman and Calcaterra, Stefano and Impelluso, Davide and Frigerio, Jacopo and Edmond, Samson and Isella, Giovanni and {Marris-Morini}, Delphine},
	year = 2026,
	month = oct,
	journal = {Optics Communications},
	volume = {616},
	pages = {133366},
	issn = {0030-4018},
	doi = {10.1016/j.optcom.2026.133366},
	urldate = {2026-09-09}
}

@article{hunt1986slownesssurface,
	title = {Slowness Surface Measurements for Zero- and Five-Degree [100]-Cut {{GaAs}}},
	author = {Hunt, W. D. and Miller, R. L. and Hunsinger, B. J.},
	year = 1986,
	month = nov,
	journal = {Journal of Applied Physics},
	volume = {60},
	number = {10},
	pages = {3532--3538},
	issn = {0021-8979, 1089-7550},
	doi = {10.1063/1.337607},
	urldate = {2024-05-06},
	langid = {english}
}

@article{jeannin2021unifieddescription,
	title = {Unified Description of Saturation and Bistability of Intersubband Transitions in the Weak and Strong Light-Matter Coupling Regimes},
	author = {Jeannin, Mathieu and Manceau, Jean-Michel and Colombelli, Raffaele},
	year = 2021,
	month = oct,
	journal = {Physical Review Letters},
	volume = {127},
	number = {18},
	pages = {187401},
	issn = {0031-9007, 1079-7114},
	doi = {10.1103/PhysRevLett.127.187401},
	urldate = {2024-12-12},
	langid = {english}
}

@article{jeannin2023lowintensity,
	title = {Low Intensity Saturation of an {{ISB}} Transition by a Mid-{{IR}} Quantum Cascade Laser},
	author = {Jeannin, Mathieu and Cosentino, Eduardo and Pirotta, Stefano and Malerba, Mario and Biasiol, Giorgio and Manceau, Jean-Michel and Colombelli, Raffaele},
	year = 2023,
	month = jun,
	journal = {Applied Physics Letters},
	volume = {122},
	number = {24},
	pages = {241107},
	issn = {0003-6951},
	doi = {10.1063/5.0153891},
	urldate = {2026-10-23}
}

@article{johnson2002perturbationtheory,
	title = {Perturbation Theory for {{Maxwell}}'s Equations with Shifting Material Boundaries},
	author = {Johnson, Steven G. and Ibanescu, M. and Skorobogatiy, M. A. and Weisberg, O. and Joannopoulos, J. D. and Fink, Y.},
	year = 2002,
	month = jun,
	journal = {Physical Review E},
	volume = {65},
	number = {6},
	pages = {066611},
	issn = {1063-651X, 1095-3787},
	doi = {10.1103/PhysRevE.65.066611},
	urldate = {2025-05-13},
	copyright = {http://link.aps.org/licenses/aps-default-license},
	langid = {english}
}

@article{jung2019homogeneousphotonic,
	title = {Homogeneous Photonic Integration of Mid-Infrared Quantum Cascade Lasers with Low-Loss Passive Waveguides on an {{InP}} Platform},
	author = {Jung, Seungyong and Palaferri, Daniele and Zhang, Kevin and Xie, Feng and Okuno, Yae and Pinzone, Christopher and Lascola, Kevin and Belkin, Mikhail A.},
	year = 2019,
	month = aug,
	journal = {Optica},
	volume = {6},
	number = {8},
	pages = {1023--1030},
	publisher = {Optica Publishing Group},
	issn = {2334-2536},
	doi = {10.1364/OPTICA.6.001023},
	urldate = {2026-09-16},
	copyright = {\copyright{} 2019 Optical Society of America},
	langid = {english}
}

@article{karnik2023highefficiencymidinfrared,
	title = {High-Efficiency Mid-Infrared {{InGaAs}}/{{InP}} Arrayed Waveguide Gratings},
	author = {Karnik, Tushar Sanjay and Dao, Khoi Phuong and Du, Qingyang and Diehl, Laurent and Pfl{\"u}gl, Christian and Vakhshoori, Daryoosh and Hu, Juejun},
	year = 2023,
	month = jan,
	journal = {Optics Express},
	volume = {31},
	number = {3},
	pages = {5056--5068},
	publisher = {Optica Publishing Group},
	issn = {1094-4087},
	doi = {10.1364/OE.480704},
	urldate = {2026-09-16},
	copyright = {\copyright{} 2023 Optica Publishing Group},
	langid = {english}
}

@article{karnik2025onchipwavelength,
	title = {On-Chip Wavelength Beam Combined {{DFB}} Quantum Cascade Laser Arrays},
	author = {Karnik, Tushar Sanjay and Diehl, Laurent and Du, Qingyang and Brian Sia, Jia Xu and Pfl{\"u}gl, Christian and Vakhshoori, Daryoosh and Hu, Juejun},
	year = 2025,
	month = apr,
	journal = {Optics Letters},
	volume = {50},
	number = {7},
	pages = {2409},
	issn = {0146-9592, 1539-4794},
	doi = {10.1364/OL.558575},
	urldate = {2026-10-23},
	langid = {english}
}

@article{khelif2010locallyresonant,
	title = {Locally Resonant Surface Acoustic Wave Band Gaps in a Two-Dimensional Phononic Crystal of Pillars on a Surface},
	author = {Khelif, Abdelkrim and Achaoui, Younes and Benchabane, Sarah and Laude, Vincent and Aoubiza, Boujamaa},
	year = 2010,
	month = jun,
	journal = {Physical Review B},
	volume = {81},
	number = {21},
	pages = {214303},
	issn = {1098-0121, 1550-235X},
	doi = {10.1103/PhysRevB.81.214303},
	urldate = {2025-01-27},
	copyright = {http://link.aps.org/licenses/aps-default-license},
	langid = {english}
}

@book{laude2015phononiccrystals,
	title = {Phononic {{Crystals}}: {{Artificial Crystals}} for {{Sonic}}, {{Acoustic}}, and {{Elastic Waves}}},
	shorttitle = {Phononic {{Crystals}}},
	author = {Laude, Vincent},
	year = 2015,
	month = sep,
	publisher = {Walter de Gruyter GmbH \& Co KG},
	googlebooks = {cCmCCgAAQBAJ},
	isbn = {978-3-11-030266-0},
	langid = {english}
}

@misc{lee2026suspendedthinfilm,
	title = {Suspended Thin-Film Lithium Niobate Modulator for Broadband Mid-Infrared Light Modulation and Frequency Comb Generation},
	author = {Lee, Chun-Ho and Ren, Xinyi and Su, Xinzhou and Lee, Wonho and Jiang, Zile and Yu, Yue and Zhou, Huibin and Zuo, Yue and Ou, Shaoyuan and Kopparapu, Reshma and Heiniger, Adam T. and Tur, Moshe and Willner, Alan E. and Chen, Zaijun and Yu, Mengjie},
	year = 2026,
	month = jan,
	number = {arXiv:2601.17385},
	eprint = {2601.17385},
	publisher = {arXiv},
	doi = {10.48550/arXiv.2601.17385},
	urldate = {2026-01-27},
	archiveprefix = {arXiv},
	langid = {english}
}

@article{li2015nanophotoniccavity,
	title = {Nanophotonic Cavity Optomechanics with Propagating Acoustic Waves at Frequencies up to 12 {{GHz}}},
	author = {Li, Huan and Tadesse, Semere A. and Liu, Qiyu and Li, Mo},
	year = 2015,
	month = sep,
	journal = {Optica},
	volume = {2},
	number = {9},
	pages = {826},
	issn = {2334-2536},
	doi = {10.1364/OPTICA.2.000826},
	urldate = {2025-05-24},
	copyright = {https://doi.org/10.1364/OA\_License\_v1\#VOR-OA},
	langid = {english}
}

@article{lin2018midinfraredintegrated,
	title = {Mid-infrared Integrated Photonics on Silicon: A Perspective},
	shorttitle = {Mid-infrared Integrated Photonics on Silicon},
	author = {Lin, Hongtao and Luo, Zhengqian and Gu, Tian and Kimerling, Lionel C. and Wada, Kazumi and Agarwal, Anu and Hu, Juejun},
	year = 2018,
	month = jan,
	journal = {Nanophotonics},
	volume = {7},
	number = {2},
	pages = {393--420},
	issn = {2192-8614, 2192-8614},
	doi = {10.1515/nanoph-2017-0085},
	urldate = {2026-10-23},
	copyright = {http://creativecommons.org/licenses/by-nc-nd/4.0/},
	langid = {english}
}

@article{lin2023realtimechirpedpulse,
	title = {Real-Time, Chirped-Pulse Heterodyne Detection at Room Temperature with 100 {{GHz}} 3-{{dB}}-Bandwidth Mid-Infrared Quantum-Well Photodetectors},
	author = {Lin, Quyang and Hakl, Michael and Lepillet, Sylvie and Li, Hua and Lampin, Jean-Fran{\c c}ois and Peytavit, Emilien and Barbieri, Stefano},
	year = 2023,
	month = dec,
	journal = {Optica},
	volume = {10},
	number = {12},
	pages = {1700},
	issn = {2334-2536},
	doi = {10.1364/OPTICA.505745},
	urldate = {2026-10-23},
	langid = {english}
}

@phdthesis{lucia2026activepassive,
	type = {Theses},
	title = {Active and {{Passive}} Devices on {{III-V}} Platform for {{Mid-Infrared Integrated Photonics}}},
	author = {Lucia, Luca},
	year = 2026,
	month = jul,
	number = {2026UPAST089},
	url = {https://theses.hal.science/tel-05760007},
	urldate = {2026-11-03},
	school = {Universit\'e Paris-Saclay}
}

@inproceedings{lucia2026fastmidinfrared,
	title = {Fast {{Mid-Infrared Devices On III-V Semiconductor Platform}}},
	booktitle = {European {{Conference}} on {{Integrated Optics}} ({{ECIO}}) 2026},
	author = {Lucia, Luca and Ngo, Gia Long and Urbain, Mathilde and Cassandra, Simone and {Huertas-Pedroche}, Javier and Pantzas, Konstantinos and Beaudoin, Gr{\'e}goire and Sagnes, Isabelle and Barbieri, Stefano and Manceau, Michel and Colombelli, Raffaele and Bousseksou, Adel},
	year = 2026,
	url = {https://opg.optica.org/abstract.cfm?uri=ECIO-2026-T1A.2},
	langid = {english}
}

@article{lucia2026midinfraredintegrated,
	title = {Mid-Infrared Integrated Resonators on a {{III}}--{{V}} Platform with {{Q}}-Factors beyond Half a Million},
	author = {Lucia, L. and Ngo, G. L. and Urbain, M. and Pirotta, S. and Pantzas, K. and Beaudoin, G. and Manceau, J. M. and Sagnes, I. and Colombelli, R. and Bousseksou, A.},
	year = 2026,
	month = mar,
	journal = {APL Photonics},
	volume = {11},
	number = {3},
	pages = {036108},
	issn = {2378-0967},
	doi = {10.1063/5.0296704},
	urldate = {2026-09-09}
}

@article{ma2013lowlosschalcogenide,
	title = {Low-Loss Chalcogenide Waveguides for Chemical Sensing in the Mid-Infrared},
	author = {Ma, Pan and Choi, Duk-Yong and Yu, Yi and Gai, Xin and Yang, Zhiyong and Debbarma, Sukanta and Madden, Steve and {Luther-Davies}, Barry},
	year = 2013,
	month = dec,
	journal = {Optics Express},
	volume = {21},
	number = {24},
	pages = {29927},
	issn = {1094-4087},
	doi = {10.1364/OE.21.029927},
	urldate = {2026-10-25},
	copyright = {https://doi.org/10.1364/OA\_License\_v1\#VOR-OA},
	langid = {english}
}

@article{malerba2024ultrafast10,
	title = {Ultrafast ({$\approx$}10 {{GHz}}) Mid-{{IR}} Modulator Based on Ultrafast Electrical Switching of the Light--Matter Coupling},
	author = {Malerba, Mario and Pirotta, Stefano and Aubin, Guy and Lucia, L. and Jeannin, M. and Manceau, J.-M. and Bousseksou, A. and Lin, Q. and Lampin, J.-F. and Peytavit, E. and Barbieri, S. and Li, L. H. and Davies, A. G. and Linfield, E. H. and Colombelli, Raffaele},
	year = 2024,
	month = jul,
	journal = {Applied Physics Letters},
	volume = {125},
	number = {4},
	pages = {041101},
	issn = {0003-6951},
	doi = {10.1063/5.0213965},
	urldate = {2026-02-15}
}

@article{mann2021ultrafastoptical,
	title = {Ultrafast Optical Switching and Power Limiting in Intersubband Polaritonic Metasurfaces},
	author = {Mann, Sander A. and Nookala, Nishant and Johnson, Samuel C. and Cotrufo, Michele and Mekawy, Ahmed and Klem, John F. and Brener, Igal and Raschke, Markus B. and Al{\`u}, Andrea and Belkin, Mikhail A.},
	year = 2021,
	month = may,
	journal = {Optica},
	volume = {8},
	number = {5},
	pages = {606--613},
	publisher = {Optica Publishing Group},
	issn = {2334-2536},
	doi = {10.1364/OPTICA.415581},
	urldate = {2026-10-23},
	copyright = {\copyright{} 2021 Optical Society of America},
	langid = {english}
}

@article{marris-morini2026germaniumbasedmidinfrared,
	title = {Germanium-{{Based Mid-Infrared Integrated Photonics}}},
	author = {{Marris-Morini}, Delphine and Mashanovich, Goran Z. and Nedeljkovic, Milos and {Alonso-Ramos}, Carlos and Vivien, Laurent and Frigerio, Jacopo and Isella, Giovanni},
	year = 2026,
	journal = {Laser \& Photonics Reviews},
	volume = {20},
	number = {14},
	pages = {e01738},
	issn = {1863-8899},
	doi = {10.1002/lpor.202501738},
	urldate = {2026-09-16},
	copyright = {\copyright{} 2026 The Author(s). Laser \& Photonics Reviews published by Wiley-VCH GmbH},
	langid = {english}
}

@article{montesinos-ballester2024lowlossburied,
	title = {Low-{{Loss Buried InGaAs}}/{{InP Integrated Waveguides}} in the {{Long-Wave Infrared}}},
	author = {{Montesinos-Ballester}, Miguel and J{\"o}chl, Elsa and Turpaud, Victor and Hillbrand, Johannes and Bertrand, Mathieu and {Marris-Morini}, Delphine and Gini, Emilio and Faist, J{\'e}r{\^o}me},
	year = 2024,
	month = jun,
	journal = {ACS Photonics},
	volume = {11},
	number = {6},
	pages = {2236--2241},
	issn = {2330-4022},
	doi = {10.1021/acsphotonics.3c01898},
	urldate = {2026-09-16}
}

@book{morgan2007surfaceacoustic,
	title = {Surface Acoustic Wave Filters: With Applications to Electronic Communications and Signal Processing},
	shorttitle = {Surface Acoustic Wave Filters},
	author = {Morgan, David P.},
	year = 2007,
	edition = {2nd ed},
	publisher = {Academic Press},
	address = {Amsterdam ; London},
	isbn = {978-0-12-372537-0},
	langid = {english},
	lccn = {TK7872.F5 M668 2007}
}

@misc{ngo2026midirsingle,
	title = {Mid-{{IR}} Single- and Dual-Electro-Optic Comb Generation with an Ultrafast Modulator},
	author = {Ngo, Gia-Long and Lucia, Luca and Pirotta, Stefano and Malerba, Mario and Li, Lianhe H. and Barbieri, Stefano and Merupo, Victor and Linfield, Edmund H. and Darqui{\'e}, Benoit and Manceau, Jean-Michel and Bousseksou, Adel and Colombelli, Raffaele},
	year = 2026,
	month = jul,
	number = {arXiv:2607.03317},
	eprint = {2607.03317},
	publisher = {arXiv},
	doi = {10.48550/arXiv.2607.03317},
	urldate = {2026-11-18},
	archiveprefix = {arXiv}
}

@article{nguyen2024integratedpin,
	title = {Integrated {{PIN}} Modulator and Photodetector Operating in the Mid-Infrared Range from 5.5 {$\mu$}m to 10 {$\mu$}m},
	author = {Nguyen, Thi Hao Nhi and Turpaud, Victor and Koompai, Natnicha and Peltier, Jonathan and Calcaterra, Stefano and Isella, Giovanni and Coudevylle, Jean-Ren{\'e} and {Alonso-Ramos}, Carlos and Vivien, Laurent and Frigerio, Jacopo and {Marris-Morini}, Delphine},
	year = 2024,
	month = apr,
	journal = {Nanophotonics},
	volume = {13},
	number = {10},
	pages = {1803--1813},
	publisher = {De Gruyter},
	issn = {2192-8614},
	doi = {10.1515/nanoph-2023-0692},
	urldate = {2026-09-16},
	copyright = {De Gruyter expressly reserves the right to use all content for commercial text and data mining within the meaning of Section 44b of the German Copyright Act.},
	langid = {english}
}

@article{perestjuk2025onemillion,
	title = {One Million Quality Factor Integrated Ring Resonators in the Mid-Infrared},
	author = {Perestjuk, Marko and Armand, R{\'e}mi and Campos, Miguel Gerardo Sandoval and Ferhat, Lamine and Reboud, Vincent and Bresson, Nicolas and Hartmann, Jean-Michel and Mathieu, Vincent and Ren, Guanghui and Boes, Andreas and Mitchell, Arnan and Monat, Christelle and Grillet, Christian},
	year = 2025,
	month = apr,
	journal = {Nanophotonics},
	volume = {14},
	number = {7},
	pages = {1009--1015},
	publisher = {De Gruyter},
	issn = {2192-8614},
	doi = {10.1515/nanoph-2024-0761},
	urldate = {2026-09-16},
	copyright = {De Gruyter expressly reserves the right to use all content for commercial text and data mining within the meaning of Section 44b of the German Copyright Act.},
	langid = {english}
}

@article{pirotta2021fastamplitude,
	title = {Fast Amplitude Modulation up to 1.5 {{GHz}} of Mid-{{IR}} Free-Space Beams at Room-Temperature},
	author = {Pirotta, Stefano and Tran, Ngoc-Linh and Jollivet, Arnaud and Biasiol, Giorgio and Crozat, Paul and Manceau, Jean-Michel and Bousseksou, Adel and Colombelli, Raffaele},
	year = 2021,
	month = feb,
	journal = {Nature Communications},
	volume = {12},
	number = {1},
	pages = {799},
	publisher = {Nature Publishing Group},
	issn = {2041-1723},
	doi = {10.1038/s41467-020-20710-2},
	urldate = {2026-09-16},
	copyright = {2021 The Author(s)},
	langid = {english}
}

@article{rembe2006laserscanningconfocal,
	title = {Laser-Scanning Confocal Vibrometer Microscope: {{Theory}} and Experiments},
	shorttitle = {Laser-Scanning Confocal Vibrometer Microscope},
	author = {Rembe, Christian and Dr{\"a}benstedt, Alexander},
	year = 2006,
	month = aug,
	journal = {Review of Scientific Instruments},
	volume = {77},
	number = {8},
	pages = {083702},
	issn = {0034-6748, 1089-7623},
	doi = {10.1063/1.2336103},
	urldate = {2025-03-11},
	langid = {english}
}

@misc{seesupplementary,
	title = {See {{Supplementary Material}} at [{{URL}} Will Be Inserted by Publisher] for a Full Sketch of the Final Device Cross-Section, Additional Considerations on Waveguide and Subwavelength Grating Designs, Information on the Heterodyne Vibrometry Setup and Additional Data on Characterization of the Acousto-Optic Efficiency for Several Devices.},
	year = 2026
}

@article{shao2019microwavetoopticalconversion,
	title = {Microwave-to-Optical Conversion Using Lithium Niobate Thin-Film Acoustic Resonators},
	author = {Shao, Linbo and Yu, Mengjie and Maity, Smarak and Sinclair, Neil and Zheng, Lu and Chia, Cleaven and {Shams-Ansari}, Amirhassan and Wang, Cheng and Zhang, Mian and Lai, Keji and Lon{\v c}ar, Marko},
	year = 2019,
	month = dec,
	journal = {Optica},
	volume = {6},
	number = {12},
	pages = {1498},
	publisher = {Optica Publishing Group},
	issn = {2334-2536},
	doi = {10.1364/optica.6.001498},
	urldate = {2025-05-23},
	copyright = {https://doi.org/10.1364/OA\_License\_v1\#VOR-OA},
	langid = {english}
}

@article{shen2017phasesensitive,
	title = {Phase Sensitive Imaging of 10 {{GHz}} Vibrations in an {{AlN}} Microdisk Resonator},
	author = {Shen, Zhen and Han, Xu and Zou, Chang-Ling and Tang, Hong X.},
	year = 2017,
	month = dec,
	journal = {Review of Scientific Instruments},
	volume = {88},
	number = {12},
	pages = {123709},
	issn = {0034-6748, 1089-7623},
	doi = {10.1063/1.4995008},
	urldate = {2025-01-10},
	langid = {english}
}

@article{simon1996couplingsurface,
	title = {Coupling of Surface Acoustic Waves to a Two-Dimensional Electron Gas},
	author = {Simon, Steven H.},
	year = 1996,
	month = nov,
	journal = {Physical Review B},
	volume = {54},
	number = {19},
	pages = {13878--13884},
	issn = {0163-1829, 1095-3795},
	doi = {10.1103/PhysRevB.54.13878},
	urldate = {2025-04-15},
	copyright = {http://link.aps.org/licenses/aps-default-license},
	langid = {english}
}

@book{snyder1984opticalwaveguide,
	title = {Optical {{Waveguide Theory}}},
	author = {Snyder, Allan W. and Love, John D.},
	year = 1984,
	publisher = {Springer US},
	address = {Boston, MA},
	doi = {10.1007/978-1-4613-2813-1},
	urldate = {2026-09-09},
	copyright = {http://www.springer.com/tdm},
	isbn = {978-0-412-24250-2 978-1-4613-2813-1},
	langid = {english}
}

@article{suk2025ultrahighqchalcogenide,
	title = {Ultra-High-{{Q}} Chalcogenide Glass on-Chip Resonators Attaining Internal Impurity-Limited Loss in the Mid-Infrared},
	author = {Suk, Daewon and Ko, Kiyoung and Wang, Rongping and Wang, Xunsi and Choi, Duk-Yong and Lee, Hansuek},
	year = 2025,
	month = jul,
	journal = {Optics Express},
	volume = {33},
	number = {15},
	pages = {32033--32040},
	publisher = {Optica Publishing Group},
	issn = {1094-4087},
	doi = {10.1364/OE.569797},
	urldate = {2026-09-16},
	copyright = {\copyright{} 2025 Optica Publishing Group},
	langid = {english}
}

@article{tadesse2014subopticalwavelength,
	title = {Sub-Optical Wavelength Acoustic Wave Modulation of Integrated Photonic Resonators at Microwave Frequencies},
	author = {Tadesse, Semere Ayalew and Li, Mo},
	year = 2014,
	month = nov,
	journal = {Nature Communications},
	volume = {5},
	number = {1},
	pages = {5402},
	issn = {2041-1723},
	doi = {10.1038/ncomms6402},
	urldate = {2025-01-15},
	langid = {english}
}

@inproceedings{teyssieux2013absolutephase,
	title = {Absolute Phase and Amplitude Mapping of {{Surface Acoustic Wave}} Fields},
	booktitle = {2013 {{Joint European Frequency}} and {{Time Forum}} \& {{International Frequency Control Symposium}} ({{EFTF}}/{{IFC}})},
	author = {Teyssieux, Damien and Baron, Thomas and Friedt, Jean-Michel and Martin, Gilles and Vairac, Pascal},
	year = 2013,
	month = jul,
	pages = {73--76},
	publisher = {IEEE},
	address = {Prague, Czech Republic},
	doi = {10.1109/EFTF-IFC.2013.6702188},
	urldate = {2023-12-08},
	isbn = {978-1-4799-0342-9},
	langid = {english}
}

@article{turpaud2024lowlosssige,
	title = {Low-Loss {{SiGe}} Waveguides for Mid-Infrared Photonics Fabricated on 200 Mm Wafers},
	author = {Turpaud, Victor and Nguyen, Thi-Hao-Nhi and Dely, Hamza and Koompai, Natnicha and Bricout, Annabelle and Hartmann, Jean-Michel and Bernier, Nicolas and Krawczyk, Julia and Lima, Gabriel and Edmond, Samson and Herth, Etienne and {Alonso-Ramos}, Carlos and Vivien, Laurent and {Marris-Morini}, Delphine},
	year = 2024,
	month = may,
	journal = {Optics Express},
	volume = {32},
	number = {10},
	pages = {17400--17408},
	publisher = {Optica Publishing Group},
	issn = {1094-4087},
	doi = {10.1364/OE.521925},
	urldate = {2026-09-16},
	copyright = {\copyright{} 2024 Optica Publishing Group},
	langid = {english}
}

@article{turpaud2024tunableonchip,
	title = {Tunable {{On-Chip Electro-Optic Frequency-Comb Generation}} at 8 \textbackslash umu \textbackslash rm m\textbackslash{} Wavelength},
	author = {Turpaud, Victor and Nguyen, Thi Hao Nhi and Koompai, Natnicha and Peltier, Jonathan and Frigerio, Jacopo and Calcaterra, Stefano and Coudevylle, Jean-Ren{\'e} and Bouville, David and {Alonso-Ramos}, Carlos and Vivien, Laurent and Isella, Giovanni and {Marris-Morini}, Delphine},
	year = 2024,
	journal = {Laser \& Photonics Reviews},
	volume = {18},
	number = {9},
	pages = {2300961},
	issn = {1863-8899},
	doi = {10.1002/lpor.202300961},
	urldate = {2026-09-16},
	copyright = {\copyright{} 2024 The Authors. Laser \& Photonics Reviews published by Wiley-VCH GmbH},
	langid = {english}
}

@article{wan2022highlyefficient,
	title = {Highly Efficient Acousto-Optic Modulation Using Nonsuspended Thin-Film Lithium Niobate-Chalcogenide Hybrid Waveguides},
	author = {Wan, Lei and Yang, Zhiqiang and Zhou, Wenfeng and Wen, Meixun and Feng, Tianhua and Zeng, Siqing and Liu, Dong and Li, Huan and Pan, Jingshun and Zhu, Ning and Liu, Weiping and Li, Zhaohui},
	year = 2022,
	month = may,
	journal = {Light: Science \& Applications},
	volume = {11},
	number = {1},
	pages = {145},
	issn = {2047-7538},
	doi = {10.1038/s41377-022-00840-6},
	urldate = {2024-11-08},
	langid = {english}
}

@article{wan2025hybridthinfilm,
	title = {Hybrid {{Thin}}-{{Film Lithium Niobate Micro}}-{{Ring Acousto}}-{{Optic Modulator}} with {{Low Half}}-{{Wave}}-{{Voltage}}-{{Length Product}}},
	author = {Wan, Lei and Huang, Jiying and Wen, Meixun and Li, Huan and Zhou, Wenfeng and Yang, Zhiqiang and Chen, Yuping and Liu, Huilong and Zeng, Siqing and Liu, Dong and Yang, Shuixian and Dai, Daoxin and Li, Zhaohui},
	year = 2025,
	month = mar,
	journal = {Laser \& Photonics Reviews},
	volume = {19},
	number = {6},
	pages = {2401832},
	issn = {1863-8880, 1863-8899},
	doi = {10.1002/lpor.202401832},
	urldate = {2026-08-13},
	langid = {english}
}

@article{wang2022monolithicintegration,
	title = {Monolithic {{Integration}} of {{Mid-Infrared Quantum Cascade Lasers}} and {{Frequency Combs}} with {{Passive Waveguides}}},
	author = {Wang, Ruijun and T{\"a}schler, Philipp and Wang, Zhixin and Gini, Emilio and Beck, Mattias and Faist, J{\'e}r{\^o}me},
	year = 2022,
	month = feb,
	journal = {ACS Photonics},
	volume = {9},
	number = {2},
	pages = {426--431},
	issn = {2330-4022},
	doi = {10.1021/acsphotonics.1c01767},
	urldate = {2026-09-16}
}

@article{xu2024unveilingefficient,
	title = {Unveiling {{Efficient Acousto-Optic Modulation}} in {{Silicon Photonic Devices}} via {{Lithium Niobate Using Transfer Printing}}},
	author = {Xu, Siyu and Liu, Weixin and Le, Xianhao and Lee, Chengkuo},
	year = 2024,
	month = sep,
	journal = {Nano Letters},
	pages = {acs.nanolett.4c03622},
	issn = {1530-6984, 1530-6992},
	doi = {10.1021/acs.nanolett.4c03622},
	urldate = {2026-02-09},
	copyright = {https://doi.org/10.15223/policy-029},
	langid = {english}
}

@article{zhang2022midinfraredmicroring,
	title = {Mid-Infrared Microring Resonators and Optical Waveguides on an {{InP}} Platform},
	author = {Zhang, Kevin and B{\"o}hm, Gerhard and Belkin, Mikhail A.},
	year = 2022,
	month = feb,
	journal = {Applied Physics Letters},
	volume = {120},
	number = {6},
	pages = {061106},
	issn = {0003-6951},
	doi = {10.1063/5.0077394},
	urldate = {2026-02-15}
}

@article{zhang2025onchip7,
	title = {On-Chip 7 {{GHz}} Acousto-Optic Modulators for Visible Wavelengths},
	author = {Zhang, Ji-Zhe and Zeng, Yu and Qin, Qing and Yang, Yuan-Hao and Tian, Zheng-Hui and Wang, Jia-Qi and Dong, Chun-Hua and Xu, Xin-Biao and Ye, Ming-Yong and Guo, Guang-Can and Zou, Chang-Ling},
	year = 2025,
	month = feb,
	journal = {Optics Express},
	volume = {33},
	number = {3},
	pages = {5562},
	issn = {1094-4087},
	doi = {10.1364/OE.540356},
	urldate = {2026-02-15},
	langid = {english}
}

@article{zhou2024electricallyinterfaced,
	title = {Electrically Interfaced {{Brillouin}}-Active Waveguide for Microwave Photonic Measurements},
	author = {Zhou, Yishu and Ruesink, Freek and Pavlovich, Margaret and Behunin, Ryan and Cheng, Haotian and Gertler, Shai and Starbuck, Andrew L. and Leenheer, Andrew J. and Pomerene, Andrew T. and Trotter, Douglas C. and Musick, Katherine M. and Gehl, Michael and Kodigala, Ashok and Eichenfield, Matt and Lentine, Anthony L. and Otterstrom, Nils and Rakich, Peter},
	year = 2024,
	month = aug,
	journal = {Nature Communications},
	volume = {15},
	number = {1},
	pages = {6796},
	publisher = {Nature Publishing Group},
	issn = {2041-1723},
	doi = {10.1038/s41467-024-51010-8},
	urldate = {2026-01-20},
	copyright = {2024 The Author(s)},
	langid = {english}
}

\onecolumngrid
\newpage

\title{Mid-Infrared Integrated Acousto-Optic Phase Modulation on GaAs/AlGaAs: Supplementary Material}

\maketitle

\onecolumngrid
\renewcommand\appendixname{}
\renewcommand{\thesection}{S\arabic{section}}
\renewcommand{\thetable}{S\arabic{table}}
\renewcommand{\thefigure}{S\arabic{figure}}
\vspace{-1cm} %

\section{Device fabrication}\label{sec:suppl_process}
\begin{figure}[ht]
	\centering
	\includegraphics{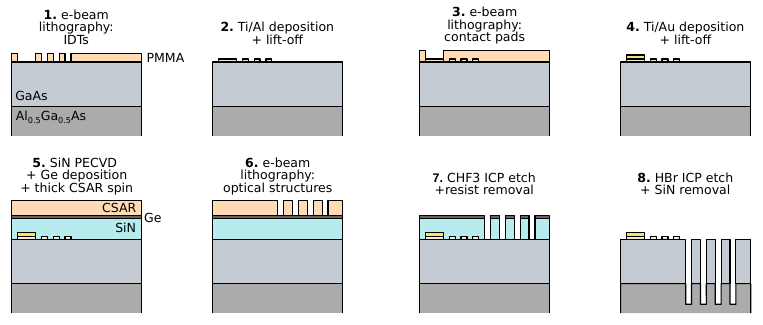}
	\caption{\label{fig:fab} Sketch of the fabrication process.}
\end{figure}

The fabrication process for the acousto-optic modulators is sketched in Fig.~\ref{fig:fab}. The devices are built on a (100)-GaAs/\ce{Al_{0.5}Ga_{0.5}As} hetero-structure grown by molecular beam epitaxy, composed of a \qty{3}{\micro m}-thick GaAs core on top of a \qty{7}{\micro m}-thick \ce{Al_{0.5}Ga_{0.5}As} cladding.

\begin{itemize}[itemsep =-3pt]
	\item We begin by defining metal marks in first e-beam lithography (EBL), metal deposition and lift-off step (not shown), which we use for re-alignment in subsequent EBL exposures. 
	\item We then define the metal IDTs in a second EBL/deposition/lift-off step of a Ti-Al \qty{10}{nm}--\qty{80}{nm} layer (1-2). Aluminum is used to minimize the electrode weight and associated IDT mass loading effects \cite{morgan2007surfaceacoustic}.
	\item A third EBL step is used to thicken the electrical contact pads with an additional Ti-Au layer (2-3), ensuring a robust electrical connection to RF probe tips.
	\item The rest of the process is dedicated to etching the optical structures, using a SiN hard mask. We first deposit a \qty{1}{\micro m}-thick layer of SiN by plasma-enhanced chemical vapor deposition, topped by \qty{40}{nm} of Ge to ensure charge evacuation during exposure as well as correct e-beam focusing, and a thick \qty{1.1}{\micro m} spin-coated layer of CSAR 62 positive resist (5). 
	\item We then define all optical structures in a single EBL exposure, including waveguide contours in double-trench geometry, mesa regions hosting the IDTs and subwavelength gratings (6).
	\item The pattern is then transfered to the SiN in a \ce{CHF3}-based inductively coupled plasma reactive ion etching (ICP-RIE) step, defining the hard mask (7). The fluorine gas also efficiently etches the thin Ge layer. Remaining resist is stripped chemically.
	\item The SiN mask is transferred into the  \mbox{III-V} heterostructure through a HBr-based ICP-RIE \cite{bouchoule2008anisotropicsmooth}, adjusting the etch time so that the \qty{3}{\micro m} GaAs core is fully removed in the sub-wavelength grating regions. Due to the higher etch rate in broader trench openings, the depth reches \qty{6.5}{\micro m} in regions surrounding the uniform waveguide sections. The remaining SiN mask is finally removed in a fluorine-based RIE (8).
\end{itemize}
Throughout the process, particular care was taken to maintain vertical sidewalls, including careful optimization of the resin profiles.

\section{Device cross section}
\begin{figure*}[!ht]
	\centering
	\makebox[\textwidth][c]{\includegraphics{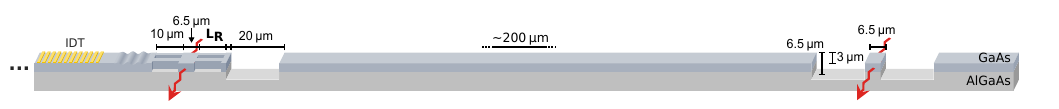}}
	\caption{\label{fig:suppl_cross_section} Cross section of the acousto-optic interaction region in the devices, showing both arms of the Mach-Zehnder interferometer.}
\end{figure*}

A sketch of the full cross section of the MZI-integrated AOM is shown in Fig.~\ref{fig:suppl_cross_section}. SAW do not reach the second arm, which is separated from the IDT by 2 \qty{6.5}{\micro m}-deep grooves and \qty{200}{\micro m} of unetched material.

\section{Optical simulations}
\begin{figure*}[!ht]
	\centering
	\makebox[\textwidth][c]{\includegraphics{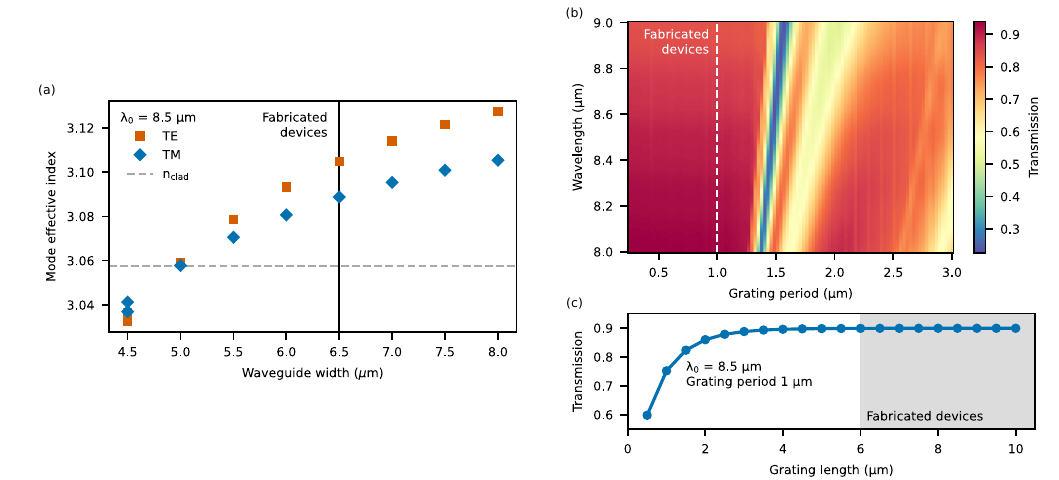}}
	\caption{\label{fig:suppl_optical_simulations} Optical FDTD simulations. (a) Passive ridge waveguide mode indices as a function of waveguide width. (b,c) Transmission through a waveguide of length \qty{50}{\micro m} surrounded by sub-wavelength grating, as a function of grating geometry.}
\end{figure*}

Design of the optical guiding structures was assisted by simulations performed with Finite-Difference Time-Domain (FDTD) software (\texttt{{Ansys\textregistered} Lumerical}). Waveguide core dimensions are chosen as a compromise between optical losses, minimized for larger cores, and acousto-optic overlap, which requires minimizing the mode spatial extension to avoid destructive acousto-optic interference in the $\lambda_\text{SAW} \approx\lambda_\text{light}$ regime. Fixing core height to \qty{3}{\micro m}, the calculated effective index for fundamental TE and TM modes is plotted as a function of waveguide width in Fig.~\ref{fig:suppl_optical_simulations}(a). The width of \qty{6.5}{\micro m} chosen for fabricated devices ensures radiative leakage in the AlGaAs cladding (dashed line) is avoided at $\lambda_0 = \qty{8.5}{\micro m}$ and below.

Simulation of the sub-wavelength grating is performed with 3D-FDTD. 
Transmission spectra through a \qty{50}{\micro m}-long waveguide surrounded by gratings are shown as a function of grating periodicity in Fig.~\ref{fig:suppl_optical_simulations}(b). In order to keep the simulations simple, the taper regions have not been included and the transition from the ridge waveguide to the grating is abrupt. 
A \qty{1.0}{\micro m} period is found to be well within the effective index guiding region for the relevant wavelength range, providing good transmission while remaining large enough to be feasibly etched over the \qty{3}{\micro m} core depth. To avoid lateral radiative leakage from the waveguide to surrounding unetched material through the grating, the same structure was simulated over a range of grating lengths (Figure~\ref{fig:suppl_optical_simulations}(c)). In fabricated samples, lengths are always kept above \qty{6}{\micro m} to keep optical losses consistent between devices.

\section{Vibrometry experimental setup}

\begin{figure*}[!ht]
	\centering
	\makebox[\textwidth][c]{\includegraphics{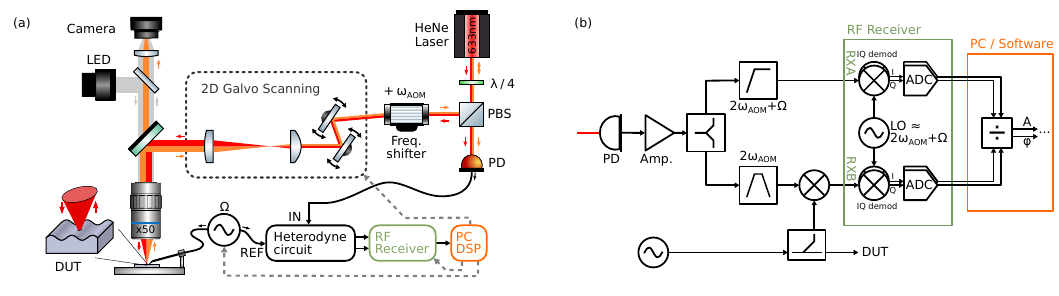}}
	\caption{\label{fig:suppl_vibrometer} Heterodyne vibrometry experimental setup. (a) Optical interferometer scheme. (b) Electronic heterodyne demodulation scheme. See text for details.}
\end{figure*}

Detection of acoustic waves is performed by reflection interferometry, in which transverse motion of a surface probed by a focused laser beam is encoded in the phase of the reflected light. The experimental setup shown in Fig.~\ref{fig:suppl_vibrometer}(a) is based on a folded heterodyne Michelson interferometer scheme (based on Ref. \cite{teyssieux2013absolutephase}), with fast scanning of the device surface enabled by a pair of motorized galvo mirrors. 

Vertically polarized light from a \qty{10}{mW} \qty{633}{nm} He-Ne laser (\texttt{LASOS LGK 7654-08}) is turned to circular polarization by a quarter-wave plate, then split evenly by a polarizing beam splitter (PBS). The horizontally-polarized beam, serving as the interferometer reference, is directly sent to a fast photodetector (\texttt{Thorlabs DET025AFC}). The vertically-polarized beam passes through an acousto-optic modulator (\texttt{G\&H AOMO 3200-125}), shifting its frequency by $\omega_\text{AOM} = \qty{200}{MHz}$. The shifted beam passes through the rotating galvo mirror pair (\texttt{Thorlabs GVSK2}, controlled by \texttt{MCC USB-234} DAQ board), before being focused on the sample by a microscope objective (\texttt{Mitutoyo M Plan Apo 50X}). A pair of relay lenses (\texttt{Thorlabs SL50-CLS2 \& TTL200MP}) placed before the objective converts rotation of the mirrors into linear translation of the laser spot on the probed surface, enabling 2D scans of devices. Reflected light takes the same path backwards, through the AOM (getting frequency shifted a second time) and the PBS onto the laser output mirror where it is reflected. When reaching the PBS again, the beam has passed through the quarter-wave plate two additional times, so it is now horizontally polarized and passes through to the detection, interfering with the reference beam. Note that since the beam returning to the laser mirror has been frequency-shifted, it does not induce a perturbation of the laser cavity. Imaging of the sample and laser spot is done using a white light LED and a lensed camera, inserted in the optical path through a shortpass dichroic beamsplitter.

Considering a transverse harmonic displacement of the probed surface $z(t) = A\cos(\Omega t+\varphi)$ and in the small amplitude regime $A\ll \lambda = \qty{633}{nm}$, the interference signal at the photodetector is composed of a carrier at the heterodyne frequency $2\omega_\text{AOM} = \qty{400}{MHz}$ and a pair of sidebands, spaced by the vibration frequency $\Omega$ \cite{teyssieux2013absolutephase}:
\begin{align}
	I(t) = I_0 + I_\text{1}&\cos(2\omega_\text{AOM}t)
	-I_1\frac{2\pi A}{\lambda}\sin(2\omega_\text{AOM}t + \Omega t + \varphi) \nonumber\\
	&+I_1\frac{2\pi A}{\lambda}\sin(2\omega_\text{AOM}t - \Omega t - \varphi).
\end{align}			
From there, the vibration amplitude is retrieved from the power ratio of any of the two sidebands to the carrier:
\begin{equation}
	A = \frac{\lambda}{2\pi}\frac{I_\text{sideband}}{I_\text{carrier}}.
\end{equation}
If detection is performed synchronously, as is the case here using the RF source exciting vibrations as a reference, the phase $\varphi$ is also retrieved from the complex sideband to carrier ratio.

The heterodyne demodulation scheme, sketched in Fig.~\ref{fig:suppl_vibrometer}(b), is based on a 2-channel software-defined radio receiver (\texttt{Ettus USRP B210}, \qty{6}{GHz}) paired with an initial analog conditioning stage. 
The photodetected signal is first amplified (\texttt{Minicircuits ZX60-14LN-S+}), then split into carrier ($2\omega_\text{AOM}$) and upper (or alternatively lower) sideband ($2\omega_\text{AOM} + \Omega$) using filters. 
The carrier is then up-converted using the driving signal to the sideband frequency, providing a reference for synchronous detection. 
The RF receiver performs simultaneous IQ downconversion and digitization of the two paths using the same local oscillator, preserving signal phase. 
The resulting complex data is streamed via USB at \qty{4}{MSa/s} per channel to a PC, where the final sideband to carrier ratio is performed by software-defined radio (\texttt{GNU Radio} \cite{blossom2004gnuradio, gnuradio}), alongside additional filtering, decimation and averaging. 
The complex ratio corresponds to amplitude and phase of the probes vibration, while an additional monitoring of the carrier amplitude provides a estimate of (relative) sample reflectivity.

Signal acquisition and scanning is automated with a \texttt{Python} (\texttt{Autolab} \cite{2026autolabprojectautolab}) interface, enabling fast 2D scans down to \qty{3}{ms} per point. Lateral resolution, limited by laser beam spot size on the sample, was measured to be $<\qty{1}{\micro m}$ (diffraction limit \qty{0.55}{\micro m} FWHM). Detection bandwidth is currently limited to $\qty{2.4}{GHz}$ by the photodetector, although a faster detector would enable detection of SAW up to \qty{6}{GHz}.

\section{Photodetector characterization \label{sec:suppl_Vigo_carac}}

\begin{figure}
	\centering
	\includegraphics{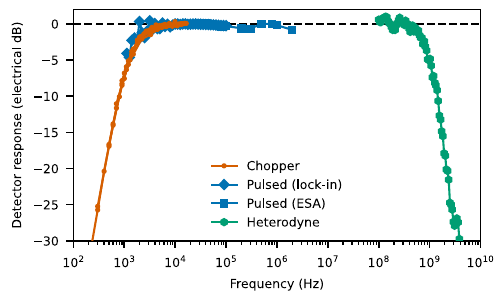}
	\caption{\label{fig:suppl_vigo_response} Measured relative frequency response of the photodetector used for optical characterization of the devices. Each dataset, measured with a different method, is adjusted so that its flat region corresponds to \qty{0}{dB} (dashed line). Measurements with a pulsed laser (blue squares) are split between data captured with a lock-in amplifier at lower frequencies and an electronic spectrum analyzer (ESA) up to \qty{2}{MHz}.}  
\end{figure}

Optical characterization of the devices relies on taking ratios of signals with frequencies ranging from optical chopping at $\qty{1070}{Hz}$ to twice the phase modulation frequency at \qty{920}{MHz}. Since these fall close to both ends of the photodetector's bandwidth (\qty{1}{kHz} to \qty{837}{MHz}, manufacturer specifications including integrated pre-amplifier), the detector's frequency response was characterized and corrected for in data analysis. Multiple measurements covering different frequency ranges are summarized in Fig.~\ref{fig:suppl_vigo_response}. Each dataset comprises a region of flat detector response, that we use to set a \qty{0}{dB} reference. 

To capture the low-frequency cut-off, we send continuous-wave laser light modulated by an optical chopper to the detector, and record (odd) harmonics of the detected square signal up $n=11$ with a lock-in amplifier. Amplitude of each harmonic as the chop frequency is varied from 100 to \qty{1500}{Hz} is measured and fitted to a 2.5-th order high-pass function. Each curve is then normalized to a unit asymptotic response (orange dotted line), and a second fit performed on the combined data from every harmonic yields a \qty{1.7}{kHz} \qty{3}{dB} electrical low-frequency cut-off (\qty{1.5}{dB} optical) over a 100 to \qty{16500}{Hz} range.

A second intermediate frequency range is attained by setting the laser to produce \qty{100}{ns} pulses, with its output attenuated then sent directly to the detector. Operating in the low duty-cycle regime, optical intensity should be proportional to the repetition rate. We measure this behavior with a lock-in amplifier or a spectrum analyzer depending on the frequency, varying the repetition rate from \qty{1.1}{kHz} to \qty{2}{MHz}, and retrieving the detector frequency response as the deviation from the expected linear behavior. The obtained curve (blue squares) matches chopper measurements at the low-end, and appears flat up to \qty{2}{MHz}.

Finally, the high-frequency cut-off (green hexagons) was covered in previously reported data \cite{marris-morini2026germaniumbasedmidinfrared}, based on heterodyne mixing of two finely controlled quantum cascade lasers \cite{hakl2021ultrafastquantumwell}. The \qty{3}{dB} electrical cut-off was measured at \qty{0.8}{GHz}. Although these measurement do not cover a continuous frequency range, we assume the response to stay flat in the MHz region.

\section{Optical characterization, additional data}	

\begin{figure}[!ht]
	\centering
	\includegraphics{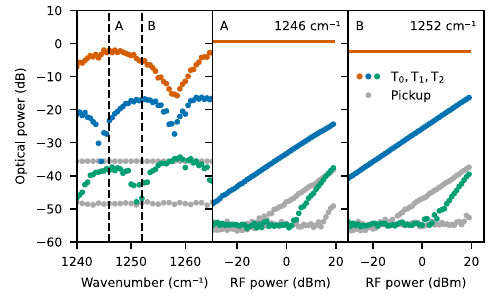}
	\caption{\label{fig:suppl_powerscans} Measured optical power at modulation harmonics as a function of injected RF power for fixed wavenumbers.  Data for harmonic $n$ is fitted to a line with slope $n/2$. The left panel shows a detail of Fig.~6 in the main text, with the two wavenumbers marked by dashed lines.}
\end{figure}		

\begin{figure*}
	\centering
	\makebox[\textwidth][c]{\includegraphics{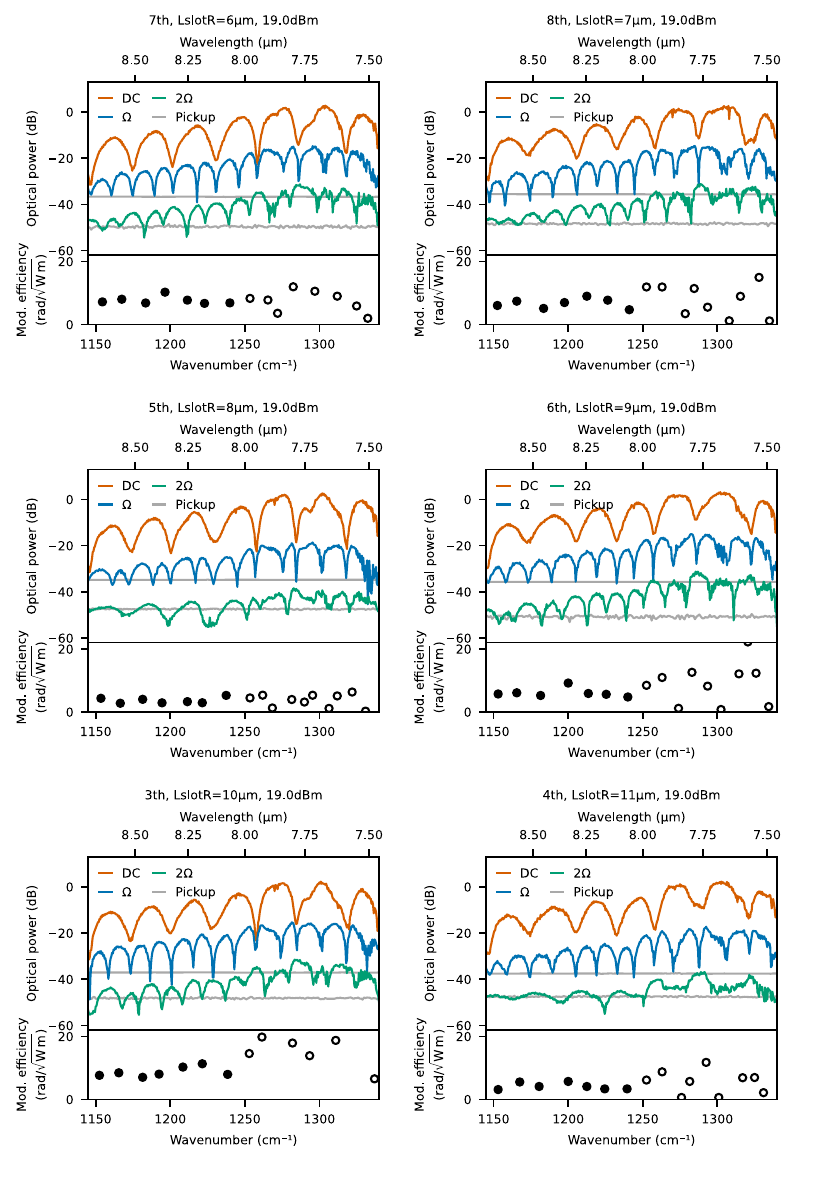}}
	\caption{\label{fig:suppl_wl_scans} Transmission spectra and modulation efficiency for six of the fabricated devices, driven at their frequency of peak efficiency. See Fig.~6 in the main text for details.}
\end{figure*}

The measured optical modulation signal is shown in Fig.~\ref{fig:suppl_powerscans} as a function of injected RF power for 2 fixed wavelengths corresponding to static phase differences of $0$ and $\pi/2$. As expected for small modulation amplitudes, optical power at the $n$-th harmonic $T_n$ scales with $(P_\text{RF})^{n/2}$. Transmission spectra equivalent to the one shown in Fig.~6 are displayed in Fig.~\ref{fig:suppl_wl_scans} for 6 other devices, along with the inferred modulation efficiencies used in Fig.~7

	\end{document}